\documentclass[useAMS,usenatbib]{mn2e}
\usepackage{amsmath}
\usepackage{graphicx}
\usepackage{epstopdf}
\usepackage{amssymb}
\usepackage{extarrows}
\usepackage{color}
\usepackage{amsmath,bm}
\usepackage{CJK}
\usepackage{hyperref}
\usepackage{forest}
\usepackage{tabularx}
\newcommand{\be}{\begin{equation}}
\newcommand{\ee}{\end{equation}}
\newcommand{\gr}{\mathrm{GR}}

\newcommand{\oct}{\mathrm{oct}}

\newcommand{\bh}{\mathrm{BH}}
\newcommand{\PN}{\mathrm{PN}}
\newcommand{\au}{\mathrm{AU}}

\newcommand{\hill}{\mathrm{Hill}}
\newcommand{\msun}{\mathrm{M_\odot}}
\newcommand{\pc}{\mathrm{pc}}

\newcommand{\kb}{\mathrm{K,B}}
\newcommand{\kc}{\mathrm{K,C}}

\newcommand{\smbh}{\mathrm{SMBH}}
\newcommand{\cm}{\mathrm{CM}}
\def\e1{e_1^2}

\title[The fate of binary stars]{The fate of close encounters between binary stars and binary supermassive black holes}
\author[Wang et al.]
{
Yi-Han Wang$^{1}$\thanks{E-mail:rosalba.perna@stonybrook.edu;\newline\quad yihan.wang.1@stonybrook.edu},
Nathan Leigh$^{2}$, Ye-Fei Yuan$^{3,4}$, Rosalba Perna$^{1}$
\\
$^{1}$ Department of Physics and Astronomy, Stony Brook University, Stony Brook, NY 11794-3800, USA\\
$^{2}$ Department of Astrophysics, American Museum of Natural History, Central Park West and 79th Street, New York, NY 10024 \\
$^{3}$ Department of Astronomy, University of Science and Technology of China, Hefei, Anhui 230026, China\\
$^{4}$ CAS Key Laboratory for Research in Galaxies and Cosmology, Hefei, Anhui 230026, China
}
\begin{document}

\pagerange{\pageref{firstpage}--\pageref{lastpage}} \pubyear{2017}

\maketitle

\label{firstpage}

\begin{abstract}
{The evolution of main sequence binaries that reside in the Galactic
  Centre can be heavily influenced by the central super massive black
  hole (SMBH). Due to these perturbative effects, the stellar binaries
  in dense environments are likely to experience mergers, collisions
  or ejections through secular and/or non-secular interactions. More
  direct interactions with the central SMBH are thought to produce
  hypervelocity stars (HVSs) and tidal disruption events (TDEs). In
  this paper, we use N-body simulations to study the dynamics of
  stellar binaries orbiting a central SMBH primary with an outer SMBH
  secondary orbiting this inner triple. The effects of the secondary
  SMBH on the event rates of HVSs, TDEs and stellar mergers are
  investigated, as a function of the SMBH-SMBH binary mass ratio. Our
  numerical experiments reveal that, relative to the isolated SMBH
  case, the TDE and HVS rates are enhanced for, respectively, the
  smallest and largest mass ratio SMBH-SMBH binaries. This suggests
  that the observed event rates of TDEs and HVSs have the potential to
  serve as a diagnostic of the mass ratio of a central SMBH-SMBH
  binary. The presence of a secondary SMBH also allows for the
  creation of hypervelocity binaries.  Observations of these systems
  could thus constrain the presence of a secondary SMBH in the
  Galactic Centre.}

\end{abstract}


\section{Introduction}

Supermassive black holes (SMBHs) ubiquitously reside at the centres of
galaxies \citep[e.g.,][]{Ho}.  Observations show that, in the inner
$\sim$ 1 parsec (pc) of our own Milky Way, the gravitational potential
is dominated by a central SMBH with mass $\sim$ 4 $\times$ 10$^6$
M$_{\odot}${  \citep[e.g.,][]{Alexander,Gillessen2017}}.  In the immediate vicinity of
the SMBH, there exists a a densely packed and complicated stellar
environment.  The relatively high stellar density at the center of the
Milky Way and the small mass of the SMBH imply that the relaxation
time of two-body interaction could be as short as a Gyr, but longer
estimates have also been quoted \citep[e.g.][]{merritt13}.

Much uncertainty persists in understanding the constituents of this
inner Keplerian-dominated region.  Some parameter space exists that
permits for the presence of an IMBH with mass 10$^3$-10$^4$
M$_{\odot}$ { \citep{Yuqingjuan,gualandris05,Gualandris2009}}.  A handful of compact
binary star systems have also been observed in the Galactic Centre
\citep{martins06}.  These bound state binaries orbiting the central
SMBH form hierarchical triple systems, and are hence subject to both
secular dynamical effects and chaotic perturbations from other objects
\citep[e.g.][]{Antonini 09,leigh16a}.  This has the potential to
stimulate a high rate of observable dynamical phenomena, such as
hypervelocity stars and tidal disruption events.

Gravitational encounters involving stellar binaries and an SMBH
operating on characteristic timescales shorter than the secular
timescale have been studied extensively: (i) When the stellar binaries
are on low angular momentum orbits, the stellar binaries are
considered to be easily broken up due to the strong tidal field of the
SMBH. As a result, one component is captured by the SMBH, while the
other is ejected at high velocity.  This is one mechanism believed to
produce hypervelocity stars (HVSs) \citep[e.g.,][]{Hills
  H,Yuqingjuan,Antonini 09}; (ii) Conversely, the stars in a binary
can be tidally disrupted by the SMBH if the relative distance is less
than the tidal disruption radius of a given binary component.  The
subsequent accretion of the stellar debris by the SMBH results in a
strong flare of electromagnetic radiation, called to a tidal
disruption event (TDE) \citep[e.g.,][]{Hills
  T,Rees,Phinney,Gezari,DoubleTDE}; (iii) Finally, such a close
interaction between the SMBH and the stellar binary may lead to a
physical collision between the two components of the binary
\citep[e.g.,][]{Ginsburg}.

If the orbital plane of the stellar binary is inclined relative to the
plane of its orbit about the SMBH by $\gtrsim 40^\circ$, the
eccentricity and inclination of the binary orbit will experience
periodic oscillations on a secular timescale, known as Lidov-Kozai
(LK) oscillations \citep[e.g.,][]{Lidov,Kozai}.  Here, the orbital
eccentricity of the binary slowly increases while the inclination
decreases, conserving angular momentum, and vice versa.  The effects
of LK oscillations include the formation of a number of exotic
astrophysical systems
\citep[e.g.,][]{Eggleton,FT,Holman,Innanen,WM,DongLai
  S,Resonance,Anderson}.  For example, the presence of a central
massive BH can accelerate the rate of black hole binary mergers due LK
oscillations, both at the centres of galaxies and globular clusters
{ \citep[e.g.,][]{Blaes,MH,Wen,AMM,Fragione2017}}.  Similarly, LK oscillations can
stimulate Type Ia supernova from white-dwarf binary mergers
\citep[e.g.,][]{Thompson,PMT} or direct collisions between
main-sequence stars and hence blue straggler formation
\citep[e.g.,][]{Katz,Kushnir,Leigh2016}.  An enhanced rate of stellar
TDEs might also be expected in the presence of a massive BH, due to
the eccentric LK mechanism alone \citep[e.g.,][]{Li TDE,Liu2017}.

In this paper, we consider stellar binaries orbiting around a central primary SMBH, 
perturbed by a distant secondary SMBH.  We study the effects of different mass ratios $q$ of the SMBH-SMBH binary 
in determining the relative rates of 
HVS production, TDE production and binary mergers.  The presence of a secondary SMBH orbiting the 
inner SMBH-binary star triple serves to perturb and/or accelerate the 
secular dynamical evolution of the inner triple system.  

The SMBH-SMBH binary, together with the stellar binary, can be decomposed into two separate hierarchical triple systems. These are the SMBH-star-star triple and the SMBH-stellar binary-SMBH triple. As anticipated, the secular dynamical processes in the inner triple are accelerated by the presence of the secondary SMBH, leading to an enhancement in the overall rate of observable astronomical events (TDEs, HVSs, mergers). 
When the inner triple has a relative orbital inclination that lies in the range $40^\circ\sim140^\circ$, 
LK oscillations occur.  Subsequently, the eccentricity of the stellar binary can be excited, leading to a merger between the two binary components.
Conversely, when LK oscillations in the outer triple are operating, 
the stellar binary orbit about the primary SMBH can be excited to very high eccentricities.  
If the periastron distance corresponding to this orbit reaches the tidal disruption radius, the binary will be disrupted, producing a TDE and/or an HVS. The mass of the secondary SMBH plays a fundamental role in this four-body system, by affecting the time scales for LK oscillations to operate.

Our paper is organized as follows.
In Section 2, we introduce the geometry of the four-body system.
Through a comparison of the relevant timescales, we characterize the parameter space
where outer LK oscillations dominate over inner LK oscillations and vice versa.
In Section 3, we describe the numerical setup and initial conditions for our N-body simulations.
In Section 4, we present the results of our N-body simulations, and explore the evolution of stellar binaries 
orbiting inside the orbit of an SMBH-SMBH binary as a function of its mass ratio.
Our conclusions are summarized in Section 5.

\section{A stellar binary orbiting an SMBH binary} 

\begin{figure}
\centering
\includegraphics[width=0.8\columnwidth]{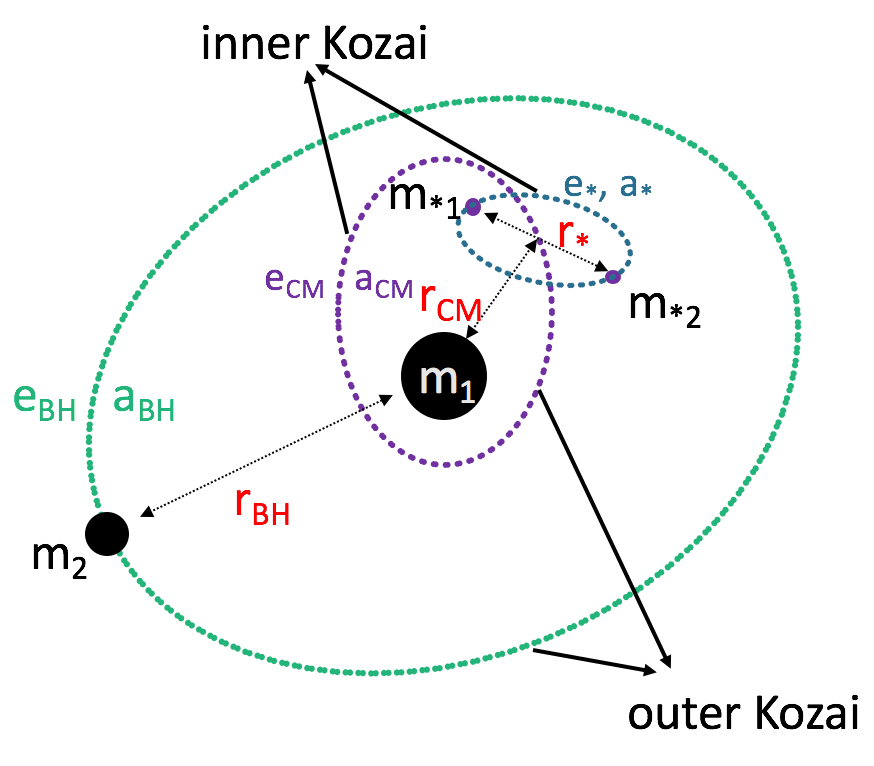}\\
\caption{Schematic illustration of the two triple systems in our four-body system. The first triple consists of the primary SMBH $m_1$ and the stellar binary $m_{*,1}$,$m_{*,2}$. The second triple consists of the primary SMBH $m_1$, 
the centre of mass of the stellar binary and the second SMBH $m_2$. 
 }
 \label{fig:1}
\end{figure}

As shown in Figure \ref{fig:1}, we label the masses of the stars by
$m_{*1}$ and $m_{*2}$ (i.e., the stellar binary components), the mass
of the primary SMBH by $m_1$, and the mass of the secondary SMBH by
$m_2$.  For the orbital parameters, we use $a_{*,\cm,\bh}$ to denote
the semi-major axes, $e_{*,\cm,\bh}$ the eccentricities and
$r_{*,\cm,\bh}$ the separations between the two components of each
binary.  Here, the subscripts ``*, CM, BH" denote, the (internal)
orbit of the stellar binary, the orbit of the stellar binary about the
primary SMBH and the orbit of the secondary SMBH about the primary
SMBH, respectively. Geometrically, the four-body system can be
decomposed in to an inner triple (SMBH-star-star) and an outer triple
(SMBH-stellar binary-SMBH).  In each triple system, if the inclination
between two orbital planes is $> 40^\circ$ and $< 140^\circ$, then the
eccentricity and inclination of the inner orbit will experience
periodic oscillations on a secular timescale, known as Lidov-Kozai
(LK) oscillations (e.g., Lidov 1962; Kozai 1962). { If the
  outer BH is on a highly inclined orbit around the binary and at a
  moderate distance, the LK oscillations breakdown. Non-secular and
  chaotic effects will be introduced into the system, which have the
  potential to lead to a rapid merger in the inner
  orbit\citep{Antonini2014}.}  In such a four-body system, the
competition between LK resonances in the inner and outer triple
configurations, { together with non-secular and chaotic
  effects, }could serve as an important catalyst for astronomically
observable events such as the production of HVSs, TDEs and stellar
mergers.

\subsection{Basic relations}
Several physical mechanisms play a fundamental role in our simulations. They are summarized in the following:
\begin{enumerate}{}{}
\item
Due to their low densities, main-sequence stars are easily tidally disrupted in the vicinity of an SMBH. Tidal disruption occurs for stars that approach the SMBH more closely than $r_{*t}$,
\be\label{eq:TDE}
r_{*t}\sim \bigg(\frac{m_{\smbh}}{m_*}\bigg)^{1/3}R_*
\ee

where $m_{\smbh}$ is the mass of the SMBH, $m_*$ the mass of the star and $R_*$ the radius of the star. In our simulations, we use the mass-radius relation $R_*/R_\odot = (m_*/m_\odot)^{0.75}$ which yields $R_*= 0.01\au$ for $m_* = 3 m_\odot$ \citep[e.g.,][]{Hansen}.

\item
Stellar binaries will be broken apart by the SMBH if the stellar
  binary enters the tidal breakup radius $r_{\mathrm{bt}}$,

\begin{equation}\label{eq:rbt}
r_{\mathrm{bt}}\sim \left(\frac{m_{\smbh}}{m_{*1}+m_{*2}}\right)^{1/3}r_*
\end{equation}
where $r_{*}$ is the separation of the two components in the stellar binary.

\item
The stellar binary remains bound to the black hole within the Hill sphere of the primary SMBH ($m_1$). The radius of the Hill sphere is

{ 
\be\label{eq:HillSphere}
R_{\hill} = a_\bh(1-e_\bh)\bigg(\frac{m_1}{3m_2}\bigg)^{1/3}
\ee
}
For stellar binaries within the Hill sphere, there is no chance either
of tidal disruption by the secondary SMBH ($m_2$) or of being ejected
from the system by the slingshot effect induced by the secondary SMBH
($m_2$). For large mass ratios of the SMBH binary, the
  stellar binaries could reside outside of the Hill sphere. For these
  stellar binaries, the hierarchical structure of the four-body system
  will be broken. The LK oscillations can no longer operate and affect
  the internal system evolution.

\item
The hierarchical condition for a stable triple system is
\be\label{eq:stable}
\varepsilon_{\oct}\equiv \frac{a_\cm}{a_\bh}\frac{e_\bh}{1-e_\bh^2}<0.1
\ee
For binaries whose orbital parameters satisfy this hierarchical condition, the LK effect can excite the binary orbit.  This could ultimately lead to decoupling of the binary followed by tidal disruptions, mergers, or HVSs. At the quadrupole level, the LK effect defines the maximum value of the excited eccentricity to be $e_{max}=\sqrt{1-\frac{5}{3}\cos^2 i_{int}}$, where $i_{int}$ is the initial inclination between the inner and outer orbital planes of the triple system. 
Combining this with Eq (\ref{eq:TDE}) and the stellar mass-radius relation, we can calculate analytically the rates of single TDEs and mergers. 

For those binaries that do not satisfy the hierarchical condition, a chaotic effect is introduced into the system. Nevertheless, scattering experiments reveal that the binary orbit can still be excited in the strong interaction region where $r_\cm \sim [0.5,2]a_\bh$\citep[e.g.,][]{liufukun}. 

\item
The hardening radius of the black hole binary is
\be
a_h = \frac{Gm_2}{4\sigma_*^2}
\ee
where $\sigma_*$ is the velocity dispersion of the stellar cusp surrounding the primary SMBH.
\end{enumerate}

\subsection{Time scales}{\label{sec:time}}

In order to identify the parameter space where the Lidov-Kozai effect becomes dominant,
we consider the different sources of apsidal precession in galactic nuclei hosting SMBHs.
For the inner triple system, the stellar binary is immersed in a dense environment.
Other perturbative effects may come into play and affect the evolution of the stellar binary, or even change the outcome of an otherwise secular interaction.

Binaries orbited by a highly inclined perturber will undergo LK oscillations.
In the inner triple, the corresponding timescale at the approximation of the quadrupole level is
\citep[e.g.,][]{Lidov,Kozai}
\begin{equation} \label{eq:LKB}
T_\kb=\frac{1}{n_*}\bigg(\frac{m_{*1}+m_{*2}}{m_1}\bigg)\bigg(\frac{a_{\cm}}{a_*}\bigg)^3(1-e_{\cm}^2)^{3/2}
\end{equation}
where $n_*=\sqrt{G(m_{*1}+m_{*1})/a_*^3}$ is the mean motion of the stellar binary.

Similarly, in the outer triple, the corresponding timescale at the quadrupole level is
\begin{equation} \label{eq:LKC}
T_\kc=\frac{1}{n_{\cm}}\bigg(\frac{m_{*1}+m_{*2}+m_1}{m_2}\bigg)\bigg(\frac{a_{\bh}}{a_{\cm}}\bigg)^3(1-e_{\bh}^2)^{3/2}
\end{equation}
where $n_{\cm}=\sqrt{G(m_{*1}+m_{*1}+m_1)/a_{\cm}^3}$ is the mean motion of the stellar binary orbiting the primary SMBH.

If the stellar binary orbital separation at pericentre is sufficiently small,
additional effects such as general relativity (GR) can dominate the tidal torque exerted by the
outer binary, suppressing the excitation of its eccentricity 
{ \citep[e.g.,][]{Blaes,Smadar 2013a,Naoz2016}}.
The timescale for precession of the argument of periapsis caused by the first order post Newtonian (PN) correction
in the inner and outer orbits is
\be\label{eq:TGRB1}
T_{\gr,*}=\frac{2\pi c^2}{3G^{3/2}}\frac{a_*^{5/2}(1-e_*^2)}{(m_{*1}+m_{*2})^{3/2}}
\ee
and
\be\label{eq:TGRB2}
T_{\gr,\cm}=\frac{2\pi c^2}{3G^{3/2}}\frac{a_{\cm}^{5/2}(1-e_{\cm}^2)}{(m_{*1}+m_{*2}+m_1)^{3/2}}
\ee
respectively. The interaction between the inner and outer binaries at 1 PN order
adds an additional term in the equation of motion. The related timescale is given by
\citep[e.g.,][]{Smadar 2013a,Clifford1,Clifford2}
\be\label{eq:TGRB3}
\begin{split}
T_{\gr,int}&=\frac{16c^2a_{\cm}^3}{9G^{3/2}a_*^{1/2}}\frac{(1-e_{\cm}^2)^{3/2}}{e_*(1-e_*^2)^{1/2}}\\
&\times\frac{(m_{*1}+m_{*2})^{3/2}}{(m_{*1}^2+m_{*1} m_{*2}+m_{*2}^2)m_1}
\end{split}
\ee

\begin{figure*}
\centering
\begin{tabular}{cc}
\includegraphics[width=\columnwidth]{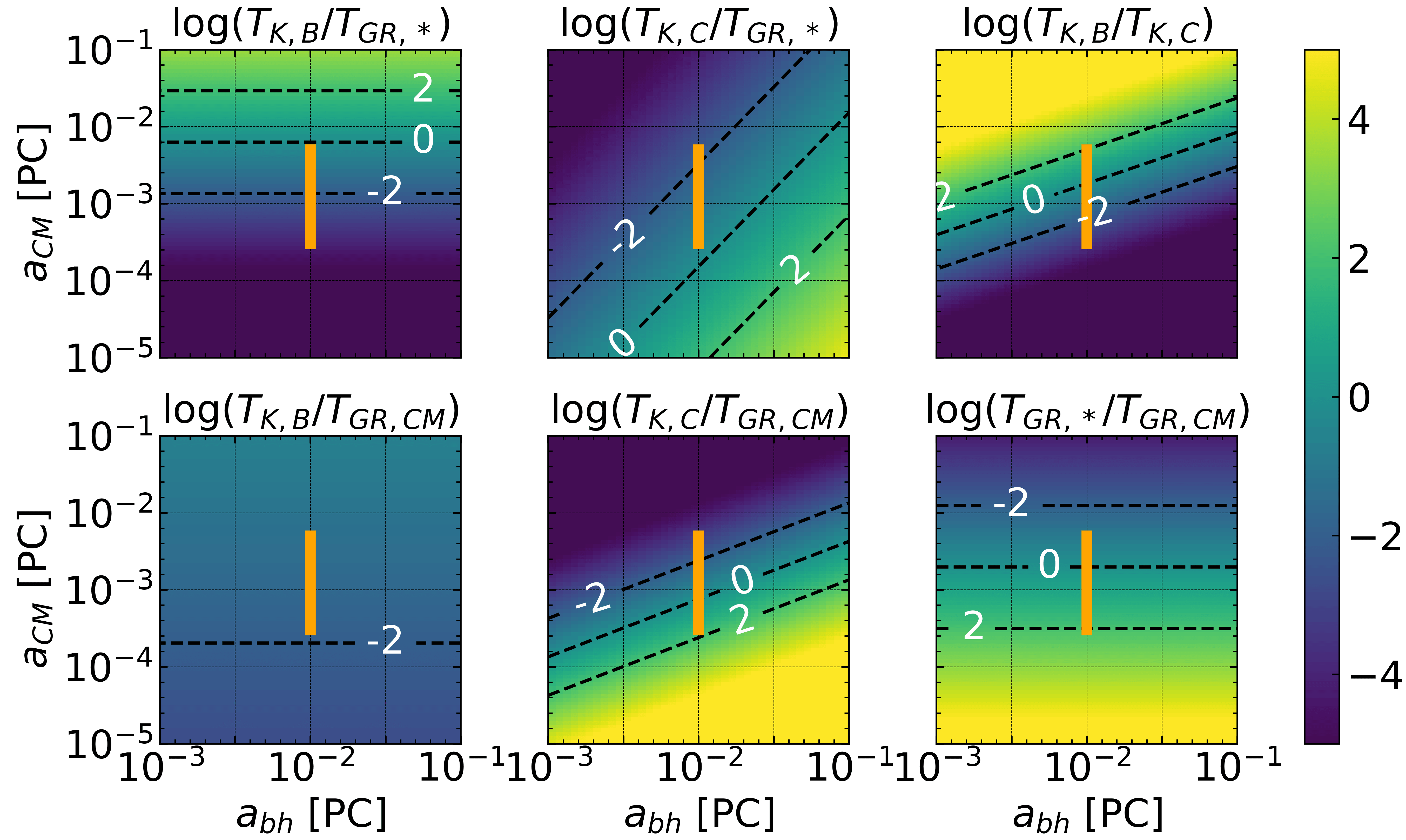}&
\includegraphics[width=\columnwidth]{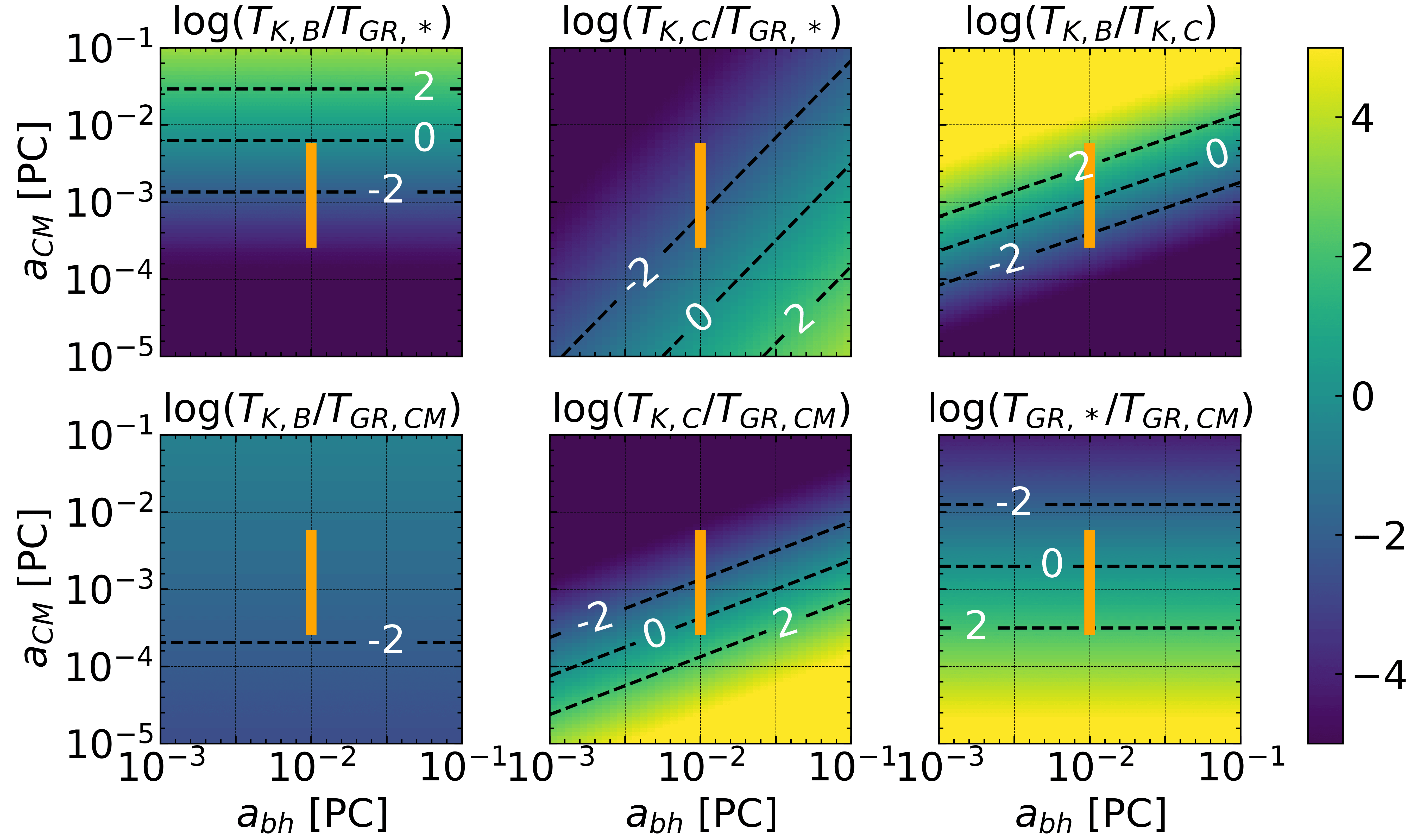} \\
\includegraphics[width=\columnwidth]{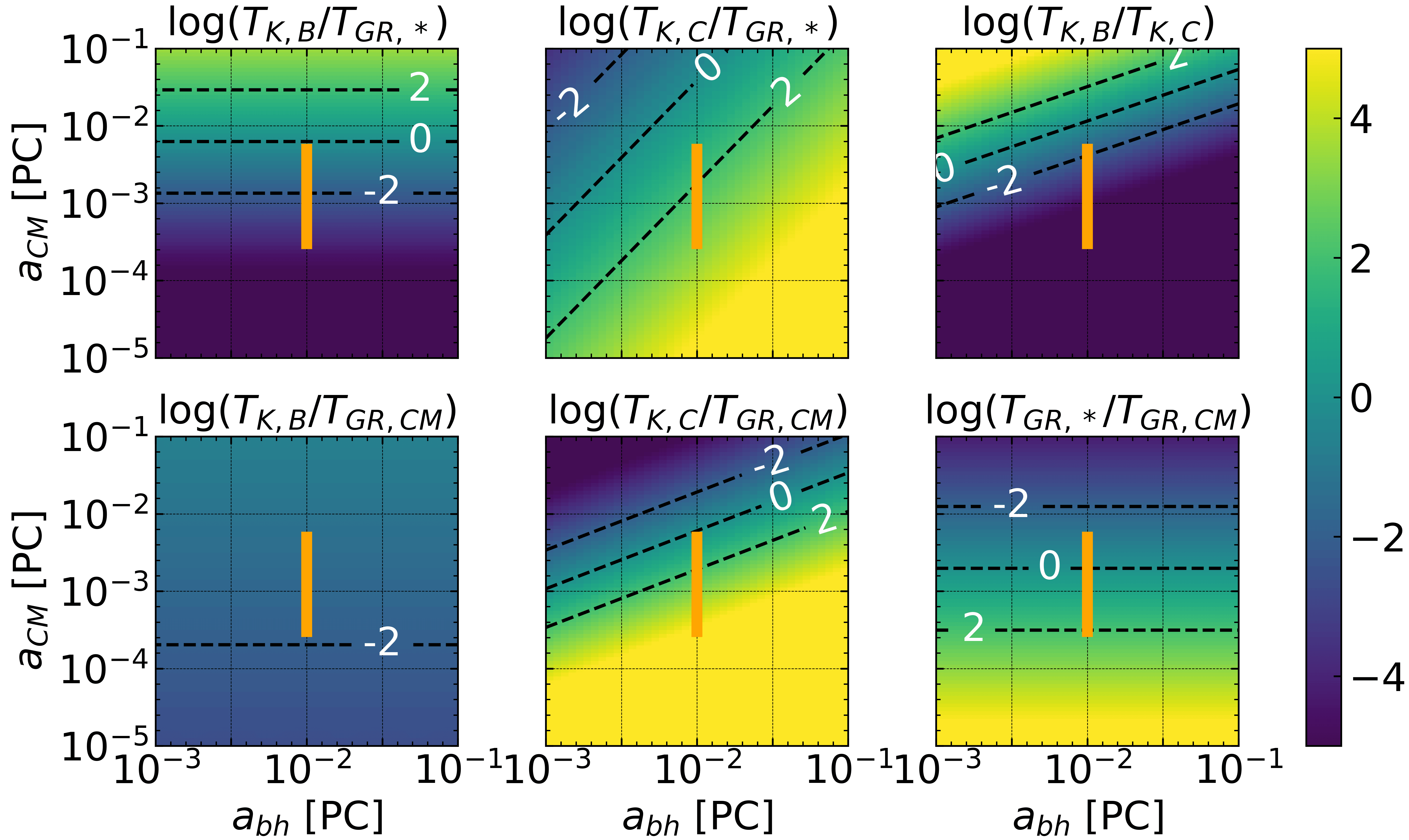}&
\includegraphics[width=\columnwidth]{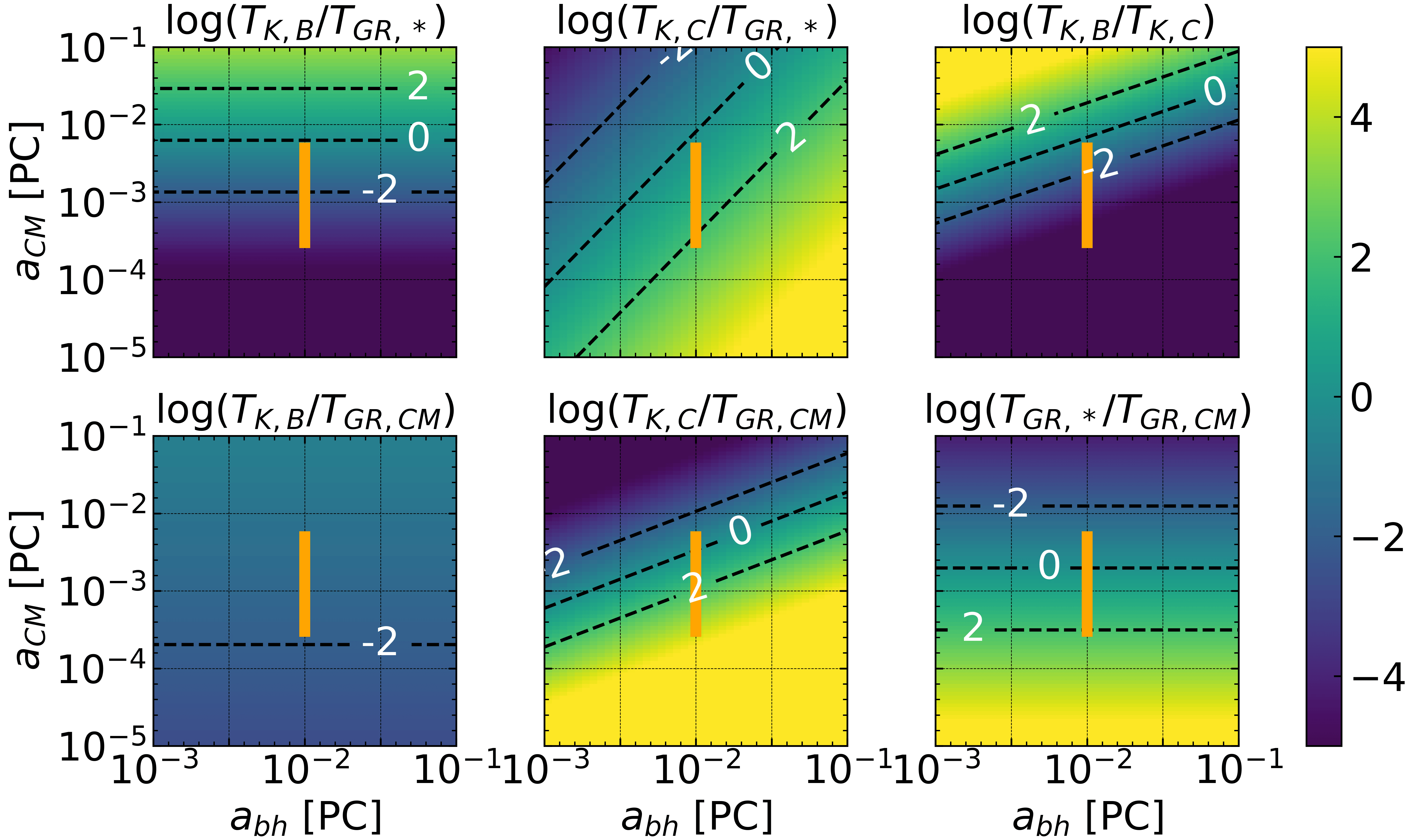} \\
\end{tabular}
\caption{Ratio of different time-scales in our four body system with
  different values of the eccentricity and mass ratio of the binary
  SMBH.  Here $T_{\kb}$, $T_{\kc}$, $T_{GR,*}$ and $T_{GR,\cm}$
  correspond to the time scales for, respectively, the inner LK
  effect, the outer LK effect, general relativistic precession of the
  stellar binary star and general relativistic precession of the
  stellar binary centre of mass. The horizontal axis shows the
  semi-major axis of the SMBH binary.  The vertical axis represents
  the distance between the stellar binary centre of mass and the SMBH.
  The axes are plotted in log scale, and the colour bar shows the
  ratio of time scales. The orange lines indicate the parameter space
  explored in our simulations.  }
\label{fig:time-scale}
\end{figure*}

Figure \ref{fig:time-scale} shows a comparison of the timescales in an illustrative example:
$m_{*1}=3 \msun$, $m_{*2}=3 \msun$ and $m_1=4\times10^6 \msun$ with different sets of mass ratio
and eccentricity of the binary SMBH.

We ran our simulations with input parameters indicated
  with orange lines in Figure \ref{fig:time-scale}. In this region,
at $a_\bh\sim0.01\pc$, it is clearly shown in all four panels that
both $T_{GR,*}$ and $T_{GR,\cm}$ are much { longer} than $T_{\kb}$. This
means that GR precession of the orbits is totally suppressed by the
inner LK oscillations at any point in the parameter space we consider.
Therefore, GR can be neglected in these simulations. The lower-middle
inset in the lower two panels indicates that GR precession of the
stellar binary centre of mass orbit dominates over the outer LK
oscillations. Thus, for small mass ratios, post-Newtonian terms must
be included in these simulations. The upper left-most inset in the
upper two panels indicates that, for large mass ratios, the inner and
outer LK oscillations compete with each other. This can serve as a
catalyst for the occurrence of a number of interesting astronomical
events, such as TDEs, HVSs and mergers. Conversely, in the lower two
panels with small mass ratios, the outer LK oscillations are
completely suppressed by the inner LK oscillations. In this case,
stellar mergers become the dominant event in our simulations.

\section{INITIAL MODELS AND NUMERICAL METHODS}
In this section, we describe the numerical method used in this paper, and present the orbital parameters and initial conditions considered for the target four-body system.

\subsection{Numerical Method}

We use $N$-body simulations to study the evolution of main sequence binaries in the Galactic Centre.  In each simulation, a stellar binary orbits the central SMBH, which is in turn orbited by a remote less-massive SMBH.  All the simulations were carried out using the code \mbox{\href{https://github.com/YihanWangAstro/NBODY}{\tt NBODY}}, developed by ourselves. In this code, we use the Position Extended Forest-Ruth Like (PEFRL) integrator with adaptive time stepping. The PEFRL algorithm is a fourth-order symplectic numerical method. Unlike other algorithms, the symplectic method conserves the constants of motion (i.e., total energy E and total angular momentum L) in Hamiltonian systems over long integration times.  We test the performance of several integrators using our $N$-body code by adopting the same set of initial conditions, performing the same simulations and comparing the results.  As expected, symplectic methods perform better than non-symplectic methods (see the Appendix A for more details).

To ensure reasonable integration times, we do not use the regular adaptive time stepping control strategy, since this requires the higher-order error estimation embedded in the PEFRL algorithm.  Instead, we adopt the time-step criterion in the cosmology code {\tt GADGET-2}\citep[e.g.,][]{Springel05}:
\begin{equation}
    \Delta t_{\mathrm{grav}} = max \bigg\{\Delta t_{\mathrm{min}},\sqrt{\frac{\eta}{|a_{\mathrm{min}}(t)|}} \bigg\}\,,
\end{equation}
where $\Delta t_{\mathrm{min}}$ is the minimal time-step allowed in
the system, to avoid an infinite shrinking of the time-step. $\eta$ is
an user-chosen accuracy parameter and $|a_{\mathrm{min}}(t)|$ gives
the minimal acceleration among all the particles in the system at any
given time $t$. This time-step criterion is only dependent on the
current status of the system.  Thus, additional integration time is
not needed to perform higher-order estimations. In our chosen problem,
given the initial total energy of the system $E(t = 0)$, the PEFRL
method controls the energy fluctuation $|\frac{E(t)-E(0)}{E(0)}|$ to
remain within the range $10^{-11}$- $10^{-13}$, assuming $\eta=0.1
R_{\odot}$.

The equation of motion is determined by the Newtonian gravitational acceleration, including post-Newtonian terms up to 2.5th order
\be
\textbf{a}_i = \textbf{a}_{\mathrm{N},i} + \textbf{a}_{\PN,i}\,,
\ee
where $a_{N,i}$ is the Newtonian acceleration imparted to the i-th particle
\be
\textbf{a}_{\mathrm{N},i} = - \sum_{j\neq i} \frac{G m_j}{r_{ij}^3}\textbf{n}
\ee
and $\textbf{n}=\textbf{r}_i-\textbf{r}_j$ above is the direction vector.\\

Post-Newtonian terms up to 2.5th order \citep[][]{Soffel1989} are included in the equation of motion in those areas of parameter space in Fig.\ref{fig:time-scale} for which the effects due to GR are important. Otherwise, we set $a_{\PN,i}$ to zero in our simulations.

\be
\textbf{a}_{\PN,i} = c^{-2}\textbf{a}_{1\PN,i}+c^{-4}\textbf{a}_{2\PN,i}+c^{-5}\textbf{a}_{2.5\PN,i}
\ee
where $\textbf{a}_{1\PN,i}$,$\textbf{a}_{2\PN,i}$ and $\textbf{a}_{2.5\PN,i}$  represent the 1st-order, 2nd-order and 2.5th-order terms, respectively\citep[e.g.,][]{PN}. Since the full equation is very long, we are not going to expand it here.
\subsection{Initial conditions and orbital parameters}\label{sec:init}

We chose the mass of the primary SMBH $m_1$ to be $4\times 10^6
M_{\odot}$, and investigated the mass ratio $q$ of the SMBH-SMBH
binary in the range $1-10^{-4}$ at [1/2,1/4,1/8,...,1/4096]. The
mass of the stars in the stellar binary are both set to be
$3M_{\odot}$.

The initial conditions of our restricted four-body problem are then
completely defined by ten configuration parameters: three for the
SMBH-SMBH binary, three for the orbit of the binary star about the
primary SMBH and four for the binary star itself.  These are:

\begin{enumerate}{}{}
\item the inclination between the SMBH binary orbit and the binary star orbit, $\theta$;
\item the longitude of the second black hole's ascending node, $l$;
\item the argument of the second black hole's pericenter, $\psi$;
\item the semi-major axis of the centre of mass of the stellar binary, $a_{\cm}$;
\item the specific angular momentum of the centre of mass of the stellar binary, $j_{\cm}$;
\item the mean anomaly of the centre of mass of the stellar binary, $M_{\cm}$;
\item the inclination of the binary star's inner orbit relative to its center of mass orbit, $\theta_*$;
\item the longitude of the stars' ascending node in the stellar binary, $l_*$;
\item the argument of the stars' pericentre in the stellar binary, $\psi_*$;
\item the mean anomaly of the stellar binary, $M_{*}$.
\end{enumerate}

For an isotropic stellar distribution, we sample $\cos\theta$ randomly
in the range [-1,1], and $l,\psi, j_{\cm}$ and $M_{\cm}$ randomly in the
range [0,2$\pi$]. $a_{\cm}$ is sampled randomly in the range
[0.03,0.5]$a_{\bh}$, while $\cos\theta_*$ is evenly sampled in the range
[-1,1].  All other phase parameters of the binary star are sampled
randomly in the range [0,2$\pi$]. The other two configuration
parameters, namely the eccentricity of the stellar binary $e_*$ (set
to 0) and the eccentricity of the SMBH binary $e_{\bh}$ (set to
[0.3,0.5,0.7] ), are fixed in each group of simulations.

We run 10,000 experiments for each group of input parameters, with an integration end-time of $5(T_{\kb}+T_{\kc})$. In each experiment, particles are deleted from the simulation upon entering either the tidal disruption radius $R_{*t}$ or the event horizon of either SMBH. Similarly, if a particle escapes to a distance $60\,a_{\bh}$ from the primary SMBH, then the particle is also deleted. A merger event occurs when the distance between two particles becomes less than the sum of their radii. We treat the merger as a completely inelastic collision, hence we delete both particles and create a new particle at the centre of mass.  All events - TDEs, mergers and escapes/HVSs - are recorded until either the maximum integration time is reached or only two particles remain in the system.

\section{Numerical exploration of the four-body system}

\subsection{Classification of the simulation results}
Four types of events are recorded in our four-body simulations. These are TDEs, breakup of stellar binaries, mergers and HVSs. Compound events are made up from these four basic events. For example, two TDEs occurring sequentially within a short time interval make a double TDE. Two HVSs without the breakup of the 
stellar binary make a double HVS. Binary disruption with a TDE makes a pure single TDE. And so on.  
In total, there are 11 kinds of compound events. 

\begin{figure}
\centering
\includegraphics[width=\columnwidth]{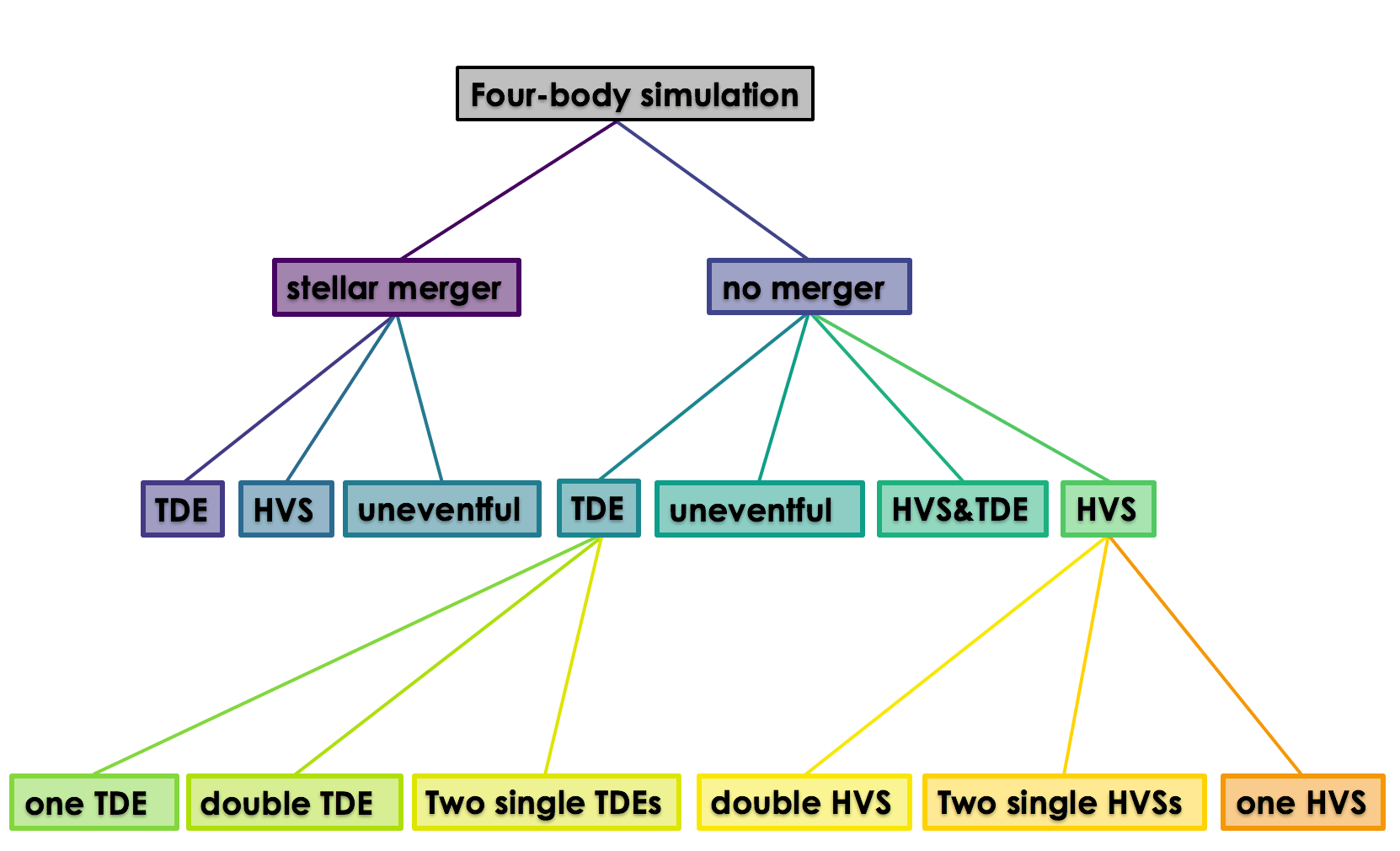}
\caption{The tree structure of different endings of the encounter of
  the stellar binaries with the binary SMBH.  They are first divided into
  two groups, with and without mergers. Once a merger occurs, the
  system becomes a three-body system. The outcomes for this three-body
  system are a TDE, an HVS or uneventful.  If no merger occurs, since
  there are two stars in the system, an extra event (HVS or TDE) is
  added to each sub-branch. At the third level of our tree, the
  TDE and HVS outcomes are both divided into either one TDE (HVS) or
  two TDEs (HVSs). At the bottom level, we assess the bound status of
  the binary star as well as the time interval between two successive
  events.  This yields the possible outcomes: a double TDE (HVS), two
  single TDEs (HVSs), or a TDE and an HVS.  }
\label{fig:eventTree}
\end{figure}

As shown in Fig.\ref{fig:eventTree}, we organize the events into a tree structure. At the top root, we perform our simulations with the initial conditions given in Section.\ref{sec:init}. The first level distinguishes between simulations with a binary merger and those without. Once a merger occurs, the four-body system changes into a three-body system. The subsequent evolution of this three-body system, including the formation of TDEs and HVSs, have been well studied. If no merger occurs, a richer set of outcomes are possible relative to the single star case. The additional star makes the second level branch of our tree possible. At the second level branch, the outcomes are TDE, HVS, HVS with TDE and uneventful. At the third level, the TDE (HVS) classification is divided into one and two TDEs (HVSs). However, due to the breakup of stellar binary, the two TDE (HVS) case can also be divided into two single TDEs (HVSs) and a double TDE (HVS).  The distinction here is the bound status of the stellar binary at the moment of occurrence of the event. The event rates of the branches at a given level or node of the tree always add to unity.

\subsubsection{Binary star mergers}
Binary star mergers are predominantly caused by orbital eccentricity
excitation from LK oscillations in the inner triple system. LK
oscillations at the quadrupole level keep the semi-major axis of the
orbit $a_*$ constant, such that an increase in the eccentricity leads
to a decrease in the pericentre distance $a_*(1-e_*)$. Once the
pericentre is sufficiently small, the two stars collide with each
other. The critical distance for this to occur is defined as the sum
of the radii of the two stars. Therefore, the merger criterion is 
\be
a_*(1-e_*) \le R_{*,1}+R_{*,2} 
\ee 
The stellar radii $R_{*,1}$ and
$R_{*,2}$ are calculated from the mass-radius relation $R_*=
(m_*/m_\odot)^{0.75}R_\odot$ \citep[e.g.,][]{Hansen}. The initial $a_{*}$ is set to
0.1$\au$ and the stellar binary has a total mass set equal to
3$m_\odot$ in our simulations. Therefore, the critical eccentricity
for merger is $e_{\mathrm{merger}}\sim 0.8$. 

\begin{figure}
\centering
\includegraphics[width=0.9\columnwidth]{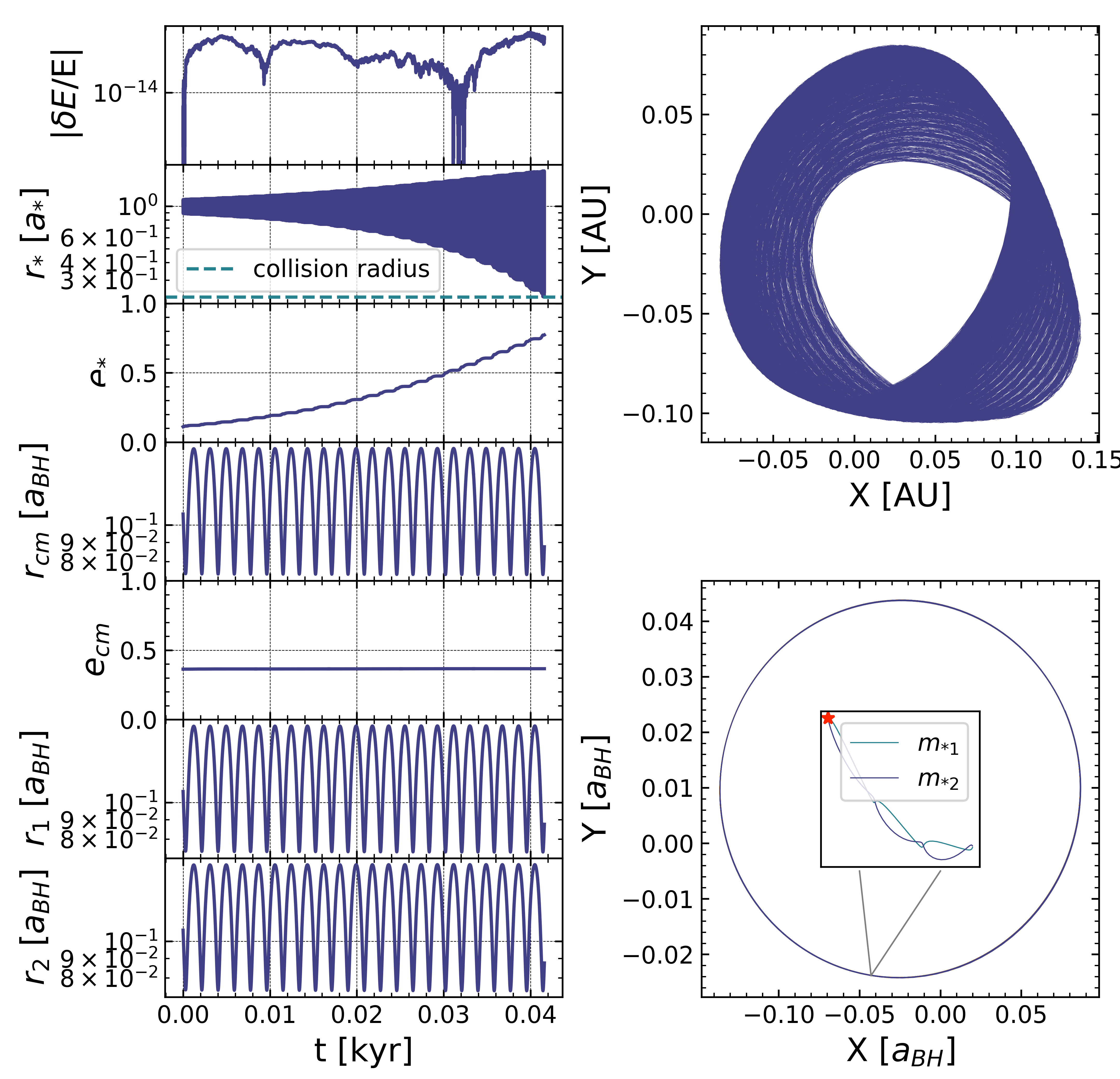}
\caption{Case study for the merging of stellar binaries.
The insets in the left panel show the total energy fluctuation and orbital parameters as a function of time, for a typical merger event. The upper right panel shows the trajectories of the stellar binary components, and the bottom right panel shows the trajectory of the stellar binary centre of mass. The initial conditions are $a_\bh = 0.01 \pc$, $e_\bh = 0.3$, $q=1/32$, $a_* =0.1\au$, $e_*=0$, $e_\cm = 0.3$, $a_\cm = 0.12 a_\bh$, with the inclinations of the inner and outer orbital planes being $i_{in} = 42.5^o$ and $i_{out} = 1.4^o$, respectively. }
\label{fig:egMerger}
\end{figure}

 { As we discuss in Sec.\ref{sec:time}, for our chosen
   initial conditions, the time scale for relativistic precession of
   the inner binary orbit $T_{GR,*}$ is much longer than $T_{\kb}$. We
   do not include dissipative tidal forces in our model. Therefore,
   the two main mechanisms for suppressing the LK oscillations and
   reducing dramatically the orbital separation in the case of mergers
   \citep{Prodan2015} cannot operate.}  Figure \ref{fig:egMerger}
 shows an illustrative example of a merger in our simulations. In this
 example, the inner inclination $i_{in}$ is larger than the critical
 angle $i_c$ required for LK oscillations, such that the inner LK
 oscillations excite the eccentricity of the stellar binary orbit
 $e_*$ until $r_*$ reaches the critical collision distance. The outer
 inclination $i_{out}$ is smaller than the critical angle, such that
 the outer LK oscillations are completely suppressed. Therefore,
 $e_\cm$ stays the same through the simulation. It is clear in this
 example that the outer LK oscillations are totally suppressed by the
 inner LK oscillations.  The outer LK oscillations can be absolutely
 suppressed if $T_\kb \ll T_\kc$, or by the initial orbital
 inclination of the outer triple falling out of the range
 $[i_c,\pi-i_c]$, where $i_c\sim 40^o$. From
 Eqs.\ref{eq:LKB},\ref{eq:LKC} and Fig.\ref{fig:time-scale}, we
 speculate that most merger events occur in the inner region (small
 initial $a_\cm$) around the SMBH where $T_\kb$ is dominant. For those
 merger events that occur in the outer region (i.e., with large
 initial $a_\cm$) where $T_\kb > T_\kc$, the outer LK oscillations are
 stifled by the small inclination $i_{out}$. In this way, the inner LK
 oscillations dominate and finally lead to a stellar merger on a
 longer timescale compared to at smaller initial $a_\cm$.

\subsubsection{Tidal Disruption}
Tidal disruption is caused by orbital eccentricity excitation in the outer triple system. The perturbation from the secondary SMBH causes the binary star to migrate in closer to the SMBH. Once the binary star reaches the stellar tidal disruption radius $r_{*t}$, a TDE occurs. However, due to chaotic perturbations to the binary star, the condition for a stable triple system in Eq. \ref{eq:stable} is not upheld in all regions around the primary SMBH. In the region where the triple system remains stable, outer LK oscillations play a fundamental role in exciting the orbital eccentricity.  Outside the stable zone, other chaotic dynamical effects can also excite the orbital eccentricity.

If the stellar binary resides in the stable three-body region, LK oscillations become the main mechanism for producing TDEs.  From Eq.\ref{eq:TDE}, the condition for a TDE is,
\be
a_\cm(1-e_\cm)\le r_{*t}
\ee
For an SMBH mass $4\times10^6 m_\odot$, $r_{*t}$ for a $3m_\odot$ star is $\sim 1.6\au=8\times 10^{-6}\pc$. In our simulations, the semi-major axis of the SMBH binary is 0.01$\pc$, and the centre of mass of the binary star is sampled evenly between $[0.03,0.5]a_\bh$. This yields the typical eccentricity for a TDE 
$e_{\mathrm{TDE}}\sim[0.99,0.999]$. 

\begin{figure}
\centering
\includegraphics[width=0.9\columnwidth]{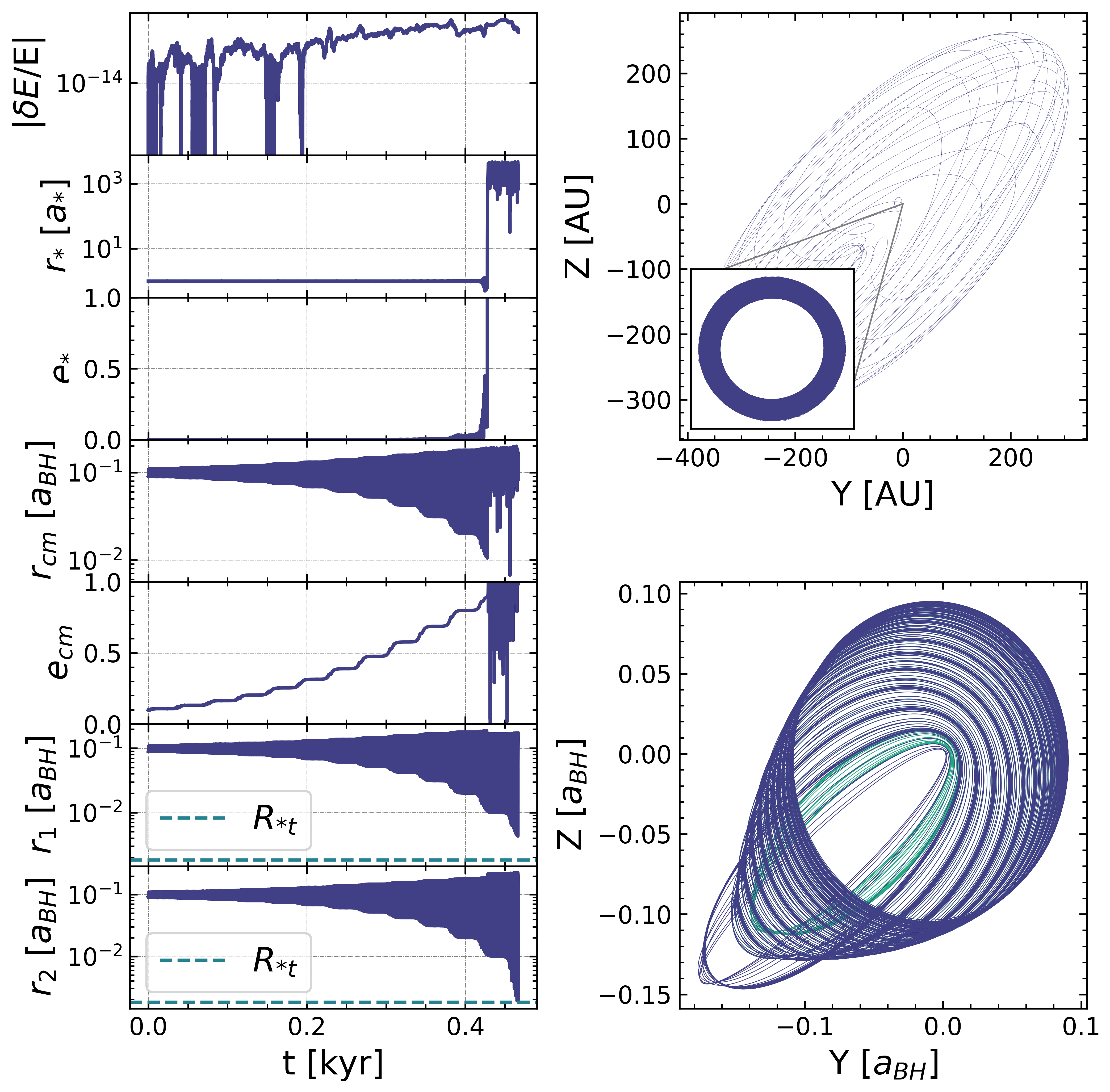}
\caption{Case study for TDE.
Same as Fig. \ref{fig:egMerger}, but the adopted initial condition is different, namely $a_\bh = 0.01 \pc$, $e_\bh = 0.5$, $q=1/2$, $a_* =0.1\au$, $e_*=0$, $e_\cm = 0.1$, $a_\cm = 0.1 a_\bh$, with the inclination of the inner and outer orbits being $i_{in} = 24.2^o$ and $i_{out} = 89.9^o$, respectively.  The stellar binary centre of mass remains in a circular orbit until the binary is disrupted.}
\label{fig:LKTDE}
\end{figure}

Figure \ref{fig:LKTDE} shows an illustrative example of a TDE from LK
oscillations. In this example, the outer inclination $i_{out}$ is
larger than the critical angle, such that the outer LK oscillations
excite the eccentricity of the binary star's centre of mass
$e_\cm$. The inner inclination $i_{in}$ is smaller than the critical
angle for LK oscillations $i_c$, such that the inner LK oscillations
are totally suppressed. The binary star maintains a perfectly circular
orbit with constant $e_* \sim 0$ until around 0.41~kyr, at which point
the stellar binary is disrupted due to the tidal force exerted by the
primary SMBH $m_1$. After this decoupling, the two stars remain in
orbit about the primary SMBH.  The outer LK oscillations continue to
excite the stars' orbits until a single TDE occurs. Unlike for binary
mergers, the inner LK oscillations need to be suppressed by the outer
LK oscillations in order for a TDE to be produced. The inner LK
oscillations can be suppressed either if $T_\kb \gg T_\kc$, or if the
initial orbital inclination of the inner triple falls outside of the
range for LK oscillations $[i_c,\pi-i_c]$. From
  Eqs.\ref{eq:LKB},\ref{eq:LKC} and Fig.\ref{fig:time-scale}, we
  conclude that $T_\kc /T_\kb$ is always smaller in the outer region
  around the SMBH. Therefore, for binary stars with initial
semi-major axis $a_\cm$ in this area, TDEs occur more frequently than
HVSs. TDE events can also occur in the inner region around the SMBH,
but only if the binary's orbit is nearly coplanar with the orbit of
the binary centre of mass about the SMBH. In this case, the inner LK
oscillations that drive stellar mergers cannot be activated due to the
low inclination $i_{in}$.  The maximum eccentricity that
can be induced by LK oscillations is constrained by the initial
inclination $e_{max}=\sqrt{1-\frac{5}{3}\cos^2 i_{int}}$.  The high
eccentricity needed for TDEs ([0.99,0.999]) requires that the orbital
plane of the binary centre of mass about the SMBH be nearly
perpendicular to the binary's orbital plane.

If Eq. \ref{eq:stable} is not satisfied, the system becomes chaotic. Chaotic effects could also excite 
the eccentricity \citep[e.g.,][]{liufukun}.   Unlike quadrupole LK oscillations, chaotic effects do not conserve the semi-major axis of the orbit.  Consequently, the subsequent evolution can be difficult to predict. However, the simulation results allow us to study chaotic excitations statistically. 

\begin{figure}
\centering
\includegraphics[width=0.9\columnwidth]{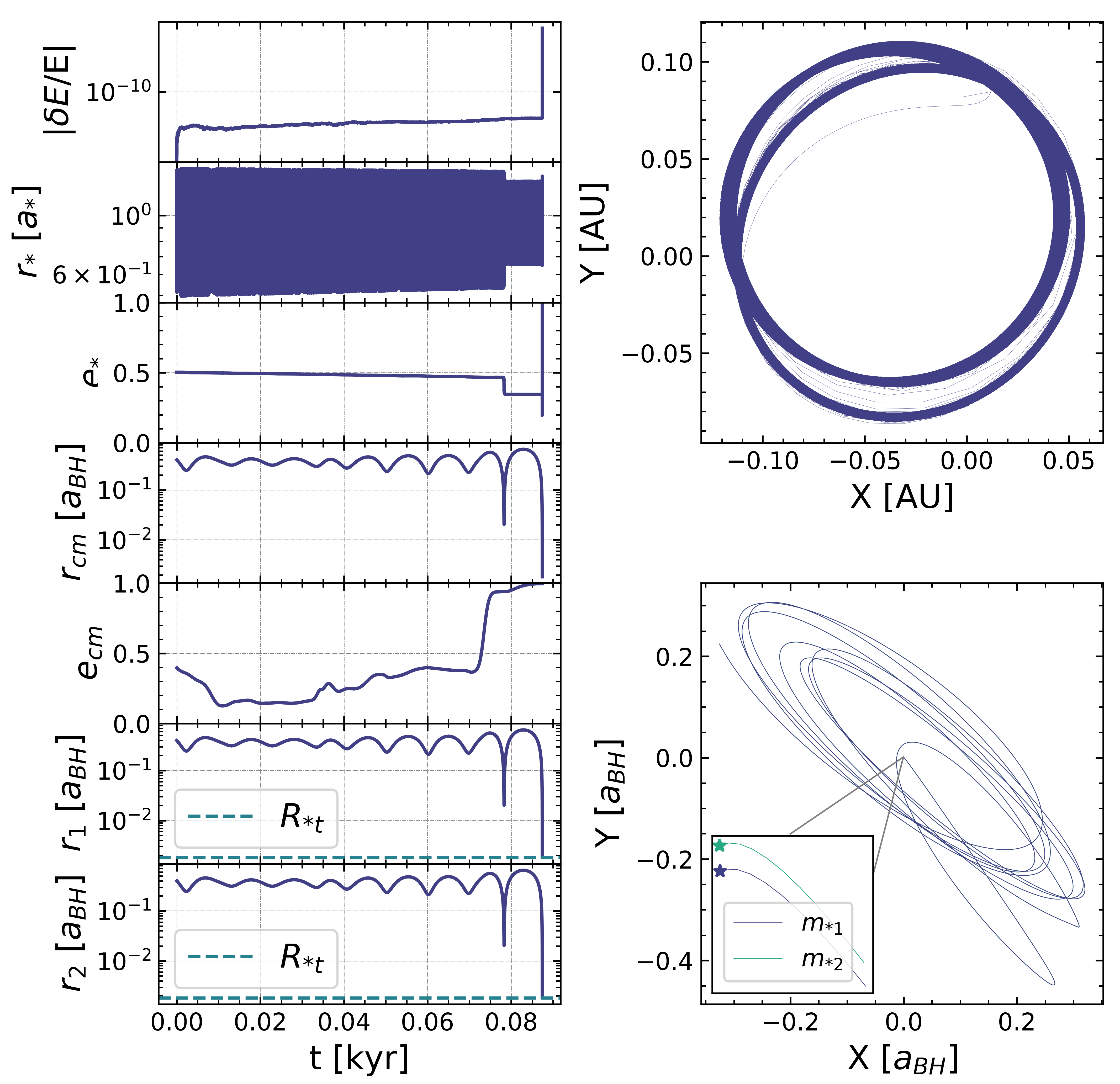}
\caption{Case study for double TDE. The initial conditions for a typical double TDE are $a_\bh = 0.01 \pc$, $e_\bh = 0.3$, $q=1/32$, $a_* =0.1\au$, $e_*=0$, $e_\cm = 0.45$, $a_\cm = 0.5 a_\bh$, with the inclination of the inner and outer orbits being $i_{in} = 5.2^o$ $i_{out} = 8.6^o$, respectively. The outer $i_{out}$ and inner $i_{in}$ inclinations are both smaller than the critical angle, such that both the inner and outer LK oscillations are suppressed. However, since $a_\cm = 0.5 a_\bh$, the system is no longer a stable hierarchical triple system, and chaotic effects excite the eccentricity $e_\cm$ in an irregular way.}
\label{fig:ChaoticTDE}
\end{figure}

Figure~\ref{fig:ChaoticTDE} shows an illustrative example of chaotic excitations. The stellar binary maintains an almost elliptical orbit at the beginning of the simulation. The first close encounter with the SMBH, which occurs around 0.08 kyr, causes the stellar binary to become more compact. At the same time, the eccentricity of the centre of mass of the stellar binary's orbit about the SMBH is suddenly excited. This excitation leads to an extremely high eccentricity, which subsequently produces a double TDE during the second close encounter. 

\subsubsection{Hypervelocity Stars}
{ 
Hypervelocity stars can be produced by the slingshot effect
\citep{2007ApJ...666L..89L}, or the interaction between a globular 
cluster and an SMBH/SMBHB\citep{Capuzzo2015,Capuzzo2016,Capuzzo2017}.  }
In our simulations, most binary
stars remain within the Hill sphere of the primary SMBH. In a stable
hierarchical triple system, LK oscillations cannot produce HVSs
because they conserve the semi-major axis of the orbit. However, if we
replace the single star in a triple system with a stellar binary, the
binary can be disrupted at the tidal breakup radius of the
  stellar binary $r_{\mathrm{bt}}$, which re-distributes energy and
angular momentum within the four-body system. This energy and angular
momentum re-allocation can produce HVSs.

\begin{figure}
\centering
\includegraphics[width=0.9\columnwidth]{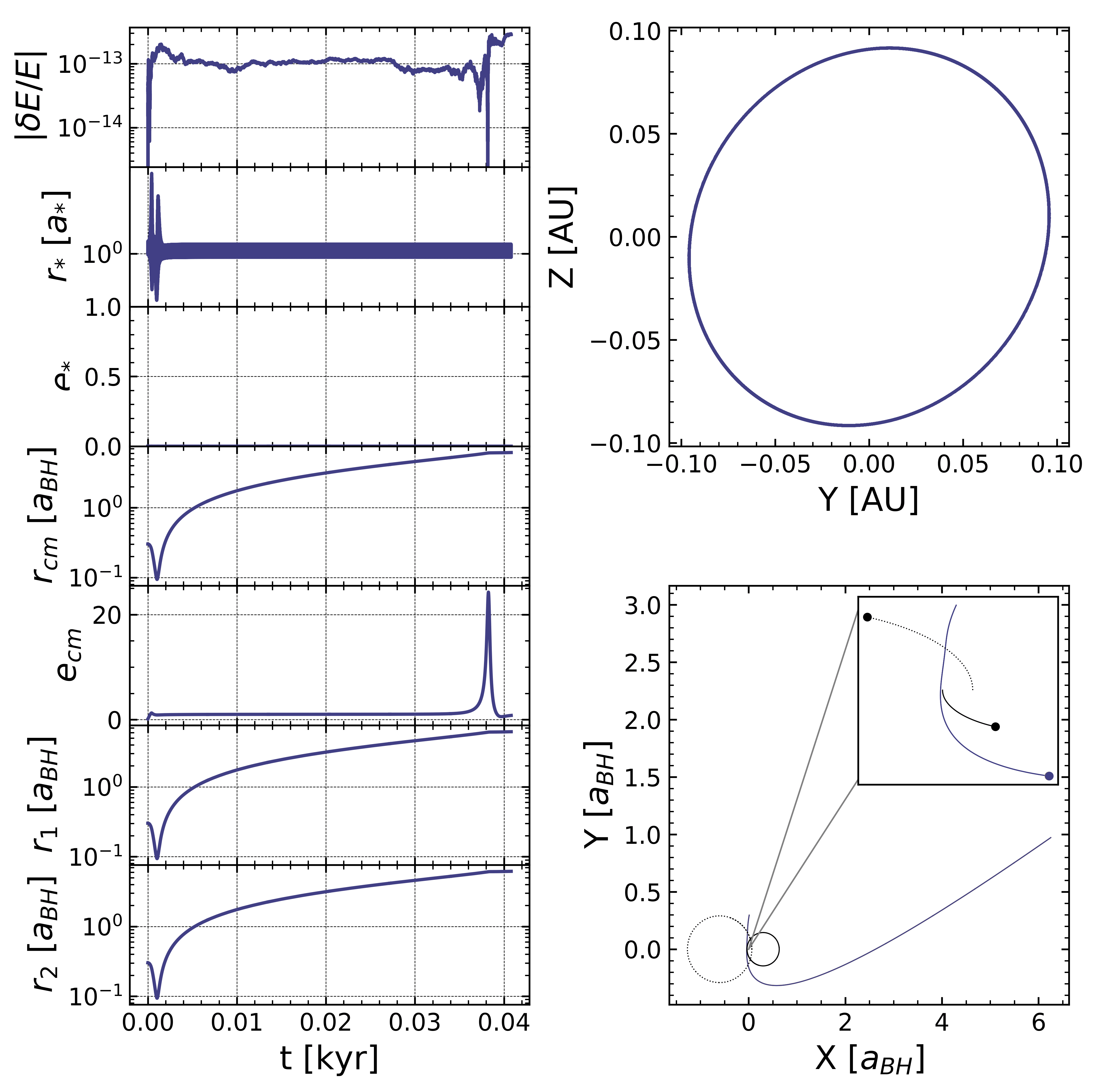}
\caption{Case study for double HVSs. The initial conditions for a typical double HVS are $a_\bh = 0.01 \pc$, $e_\bh = 0.9$, $q=1/2$, $a_* =0.1\au$, $e_*=0$, $e_\cm = 0.11$, $a_\cm = 0.34 a_\bh$, with the inclination of the inner and outer orbits being $i_{in} = 1.4^o$ $i_{out} = 67.3^o$, respectively.  After the ejection, the stellar binary recedes from the SMBH binary system. Consequently, the eccentricity $e_\bh$, which is defined as the orbital eccentricity around the primary SMBH $m_1$, diverges to infinity.}
\label{fig:HVS}
\end{figure}

Figure \ref{fig:HVS} shows an illustrative example of a double
HVS. The inner $i_{in}$ inclination is smaller than the critical
angle, such that the inner LK oscillations are suppressed. The outer
$i_{out}$ inclination is larger than the critical angle. Therefore,
the outer LK oscillations should result in an excitation of the orbit
of the centre of mass of the stellar binary. However, since $a_\cm =
0.34 a_\bh$, $q=1/2$ and $e_\bh = 0.9$, the secondary SMBH gets very
close to the stellar binary, leading instead to an ejection event.
The distance to the SMBH of the stellar binary is insufficiently small
for it to be disrupted after a slight oscillation, and a double HVS
occurs instead. Most HVSs are the result of a strong perturbation from
the secondary SMBH, which occurs most frequently for larger mass
ratios $q$ and higher eccentricities $e_\bh$. Such a strong
perturbation can even result in a double HVS event.

\subsection{Event rates for different SMBH binary mass ratios}
LK oscillations in the inner and outer triple, together with non-secular effects and chaotic
effects, influence the orbital evolution of the main sequence
binary. {  The efficiency of the inner and outer LK oscillations depends on the relative ratio of their respective time scales.
As Eq. \ref{eq:LKB} and \ref{eq:LKC} shown, the relative time scales of the inner and outer LK oscillations are sensitive to the ratios  $a_\cm/a_*$ and $a_\bh/a_\cm$. In our simulations, we fix the semi major axis of the stellar binary to reduce the free parameters in our model.  Regardless, how a? affects the outcome is readily apparent from our simulations. When the inner LK oscillations dominate, the fraction of stellar mergers becomes large (as shown in Fig.\ref{fig:stellarmerger}). 
 In this case, the choice of binary mass and semi-major axis will affect the merger rate significantly.  If the semi-major axis of the SMBH-SMBH binary and that of the stellar binary's orbit about the primary SMBH are fixed, compact stellar binaries are harder to merger.  This is because smaller values for $a_*$ yield longer time scales for the inner LK oscillations to operate. On the other hand, when the outer LK oscillations dominate, the choice of binary mass and semi-major axis do not influence the TDE and HVS event rates significantly. Here, the stellar binary can effectively be regarded as a single particle. 
 
The presence of a background gravitational potential could also affect the relative event rates.  In particular, in the Galactic Centre, dim low-mass stars are not detectable, and it is not known if any contribute to the local potential in the immediate vicinity of the primary SMBH.  Any contribution to the potential from gas and/or dust is also not included in our calculations.  Naively, however, we expect these contributions to have a negligible effect on the results reported in this paper.  Scattering experiments under the presence of a background potential were recently performed by \citep{Ryu17a}, and subsequently applied to the production of runaway and hypervelocity stars  \citep{Ryu17b}.It was shown that, unless the potential is very deep, the outcome of the scattering experiment is not significantly affected.}

\begin{figure*}
\centering
\includegraphics[width=\columnwidth]{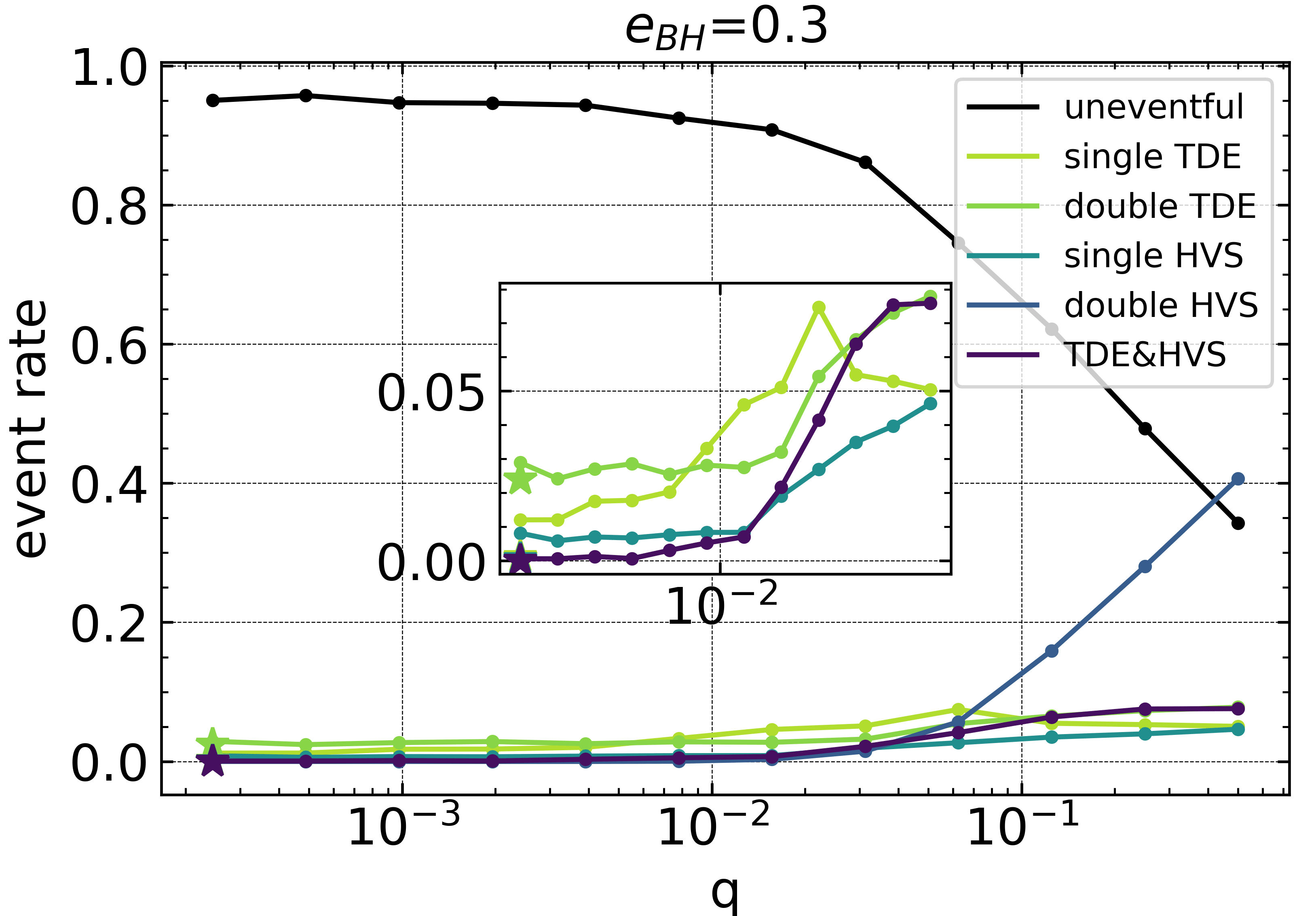}
\includegraphics[width=\columnwidth]{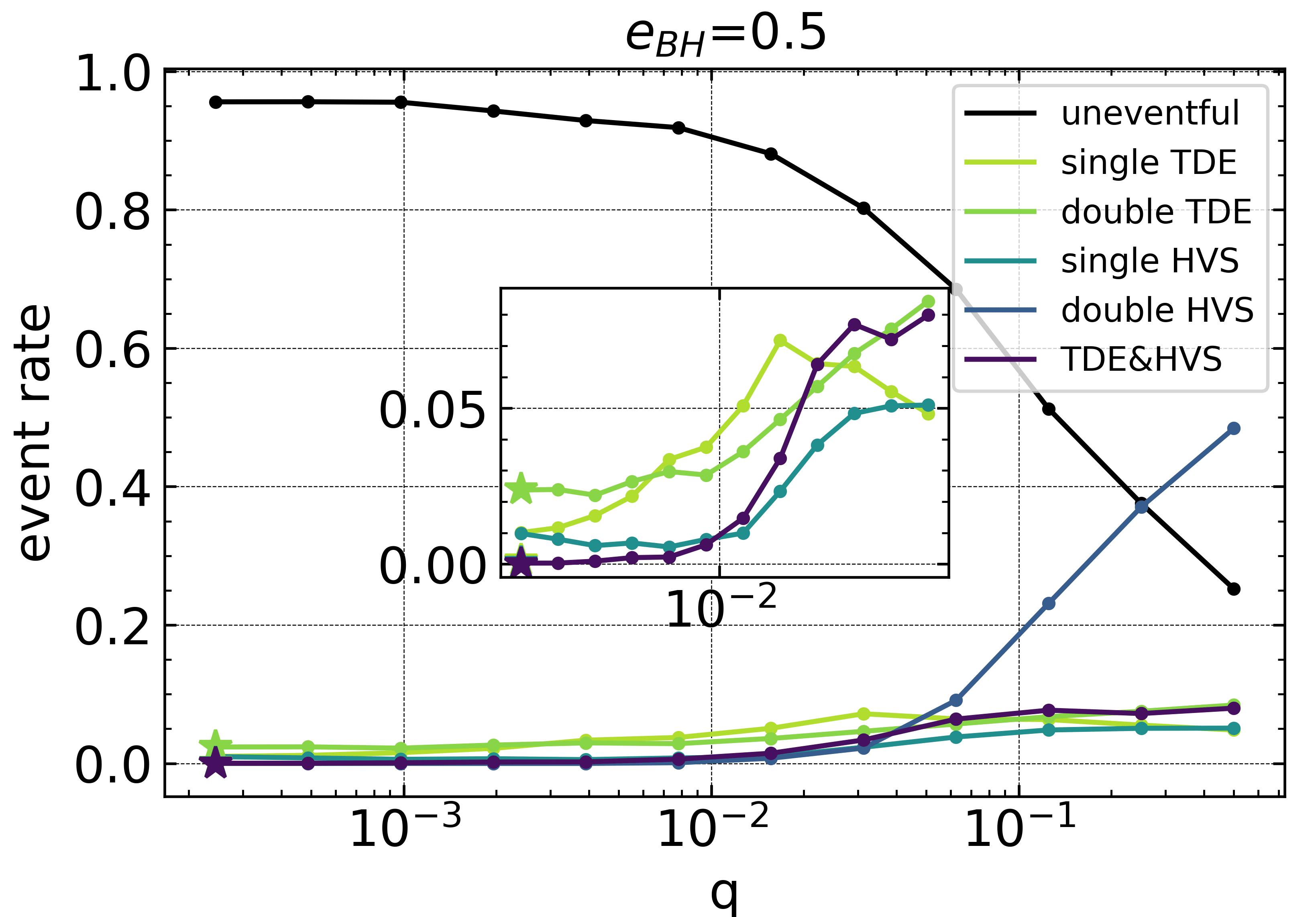} \\
\includegraphics[width=\columnwidth]{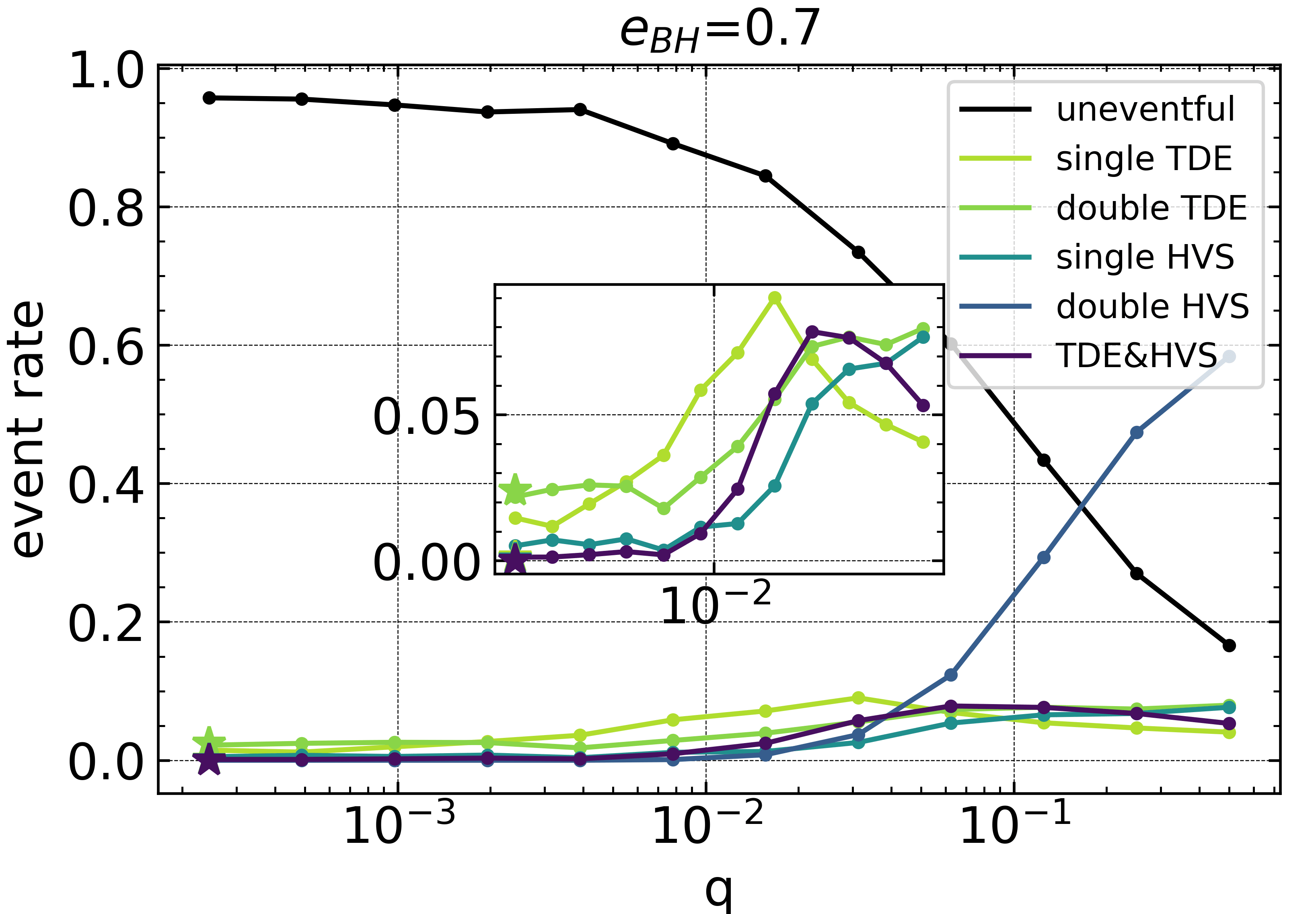}
\includegraphics[width=\columnwidth]{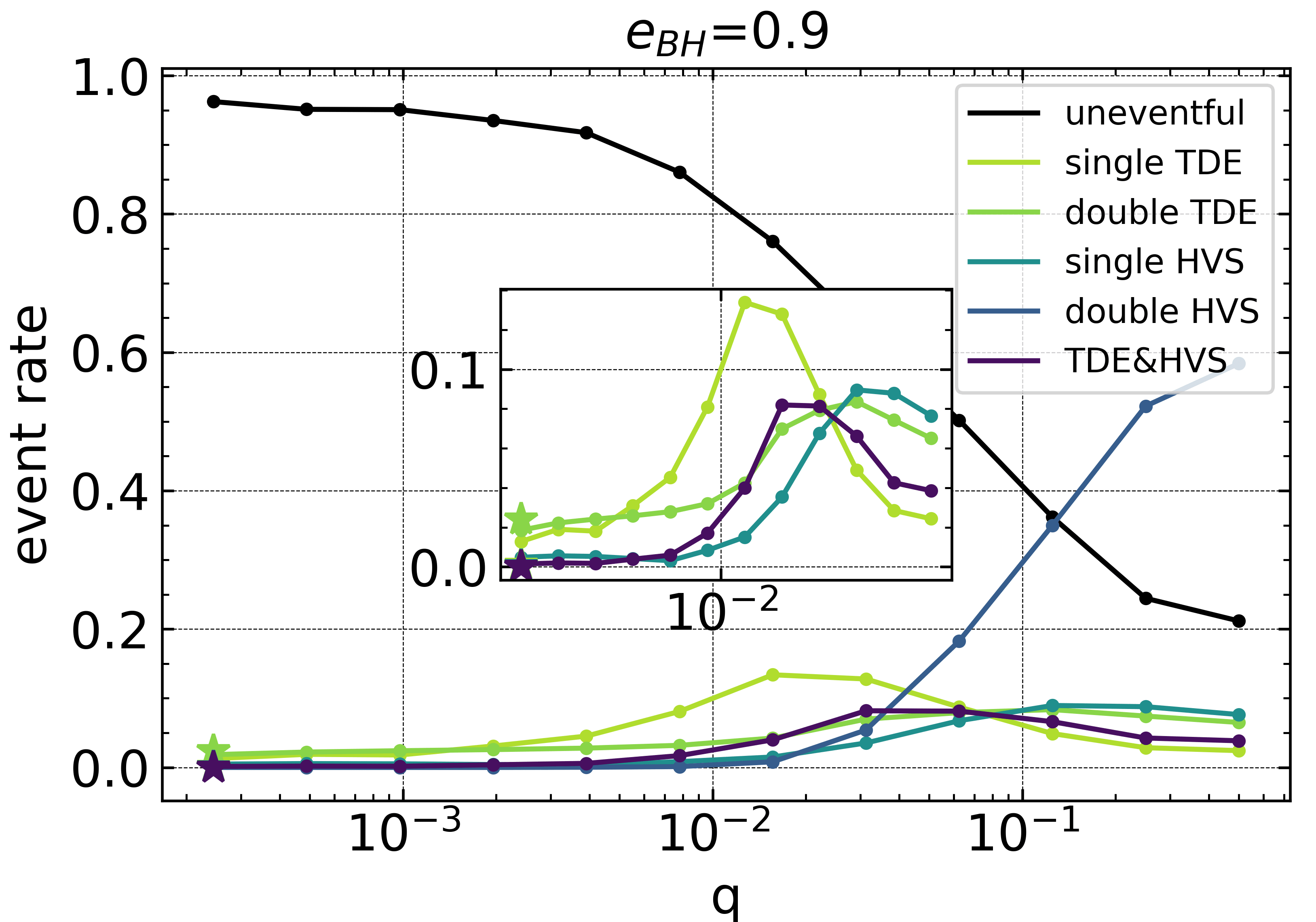} \\
\caption{Event rates as a function of mass ratio of binary SMBH. All events are shown in Fig. \ref{fig:eventTree}. These events are:  uneventful, single TDE, double TDE, single HVS, double HVS and TDE with HVS. The HVS rates continually increase after the activation threshold near $q \sim 10^{-2}$. Larger eccentricities $e_\bh$ accelerate the rate of increase, due to stronger perturbations at pericentre.  The maximum TDE rates occur between $q \sim 10^{-2}$ and $q \sim 10^{-1}$.}
\label{fig:allrates}
\end{figure*}

 To quantify the significance of the secondary SMBH in deciding
each outcome, we vary the mass ratio of the SMBH binary. Figure
\ref{fig:allrates} depicts the rates of different events as a function
of the mass ratio $q$, for different eccentricities $e_\bh$. 

As the mass ratio increases, so do the total event rates.  This is
especially true for the double HVS rates.  The TDE rates form a peak
between $q=10^{-2}$ and $q=10^{-1}$.   The 'star' at the
  left end of each curve shows the event rates for a single SMBH
  ($q=0$) { obtained from three-body simulations}. The event rates for the SMBH-SMBH binary converge to the
  single SMBH case when $q \rightarrow 0$, except for single TDEs and
  single HVSs.  The single TDEs (HVSs) come from the decoupling of the
  stellar binary, which is sensitive to the mass of the secondary
  SMBH. For the mass ratio $1/4096$, the mass of the secondary SMBH is
  still not small enough to neglect the effect on decoupling due to
  the secondary SMBH. Overall, the event rates for an SMBH-SMBH
binary are always different than those obtained for a single SMBH.
Consequently, the observed relative event rates can be used to
constrain the possible presence of a binary SMBH companion orbiting
the central primary SMBH in the Galactic Centre.  We will return to
this interesting possibility in Section~\ref{discussion}.

\begin{figure}
\centering
\includegraphics[width=0.9\columnwidth]{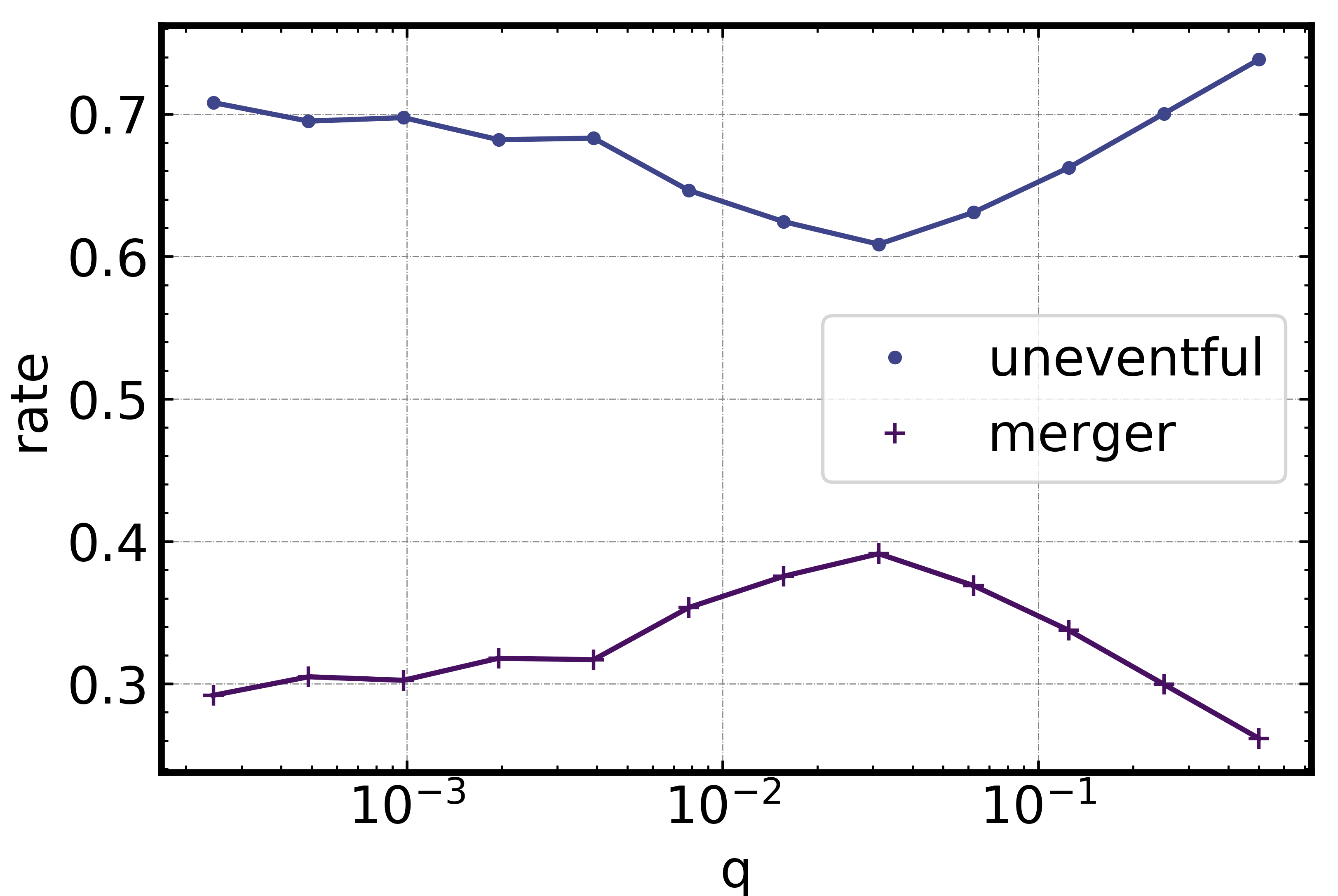} \\
\includegraphics[width=\columnwidth]{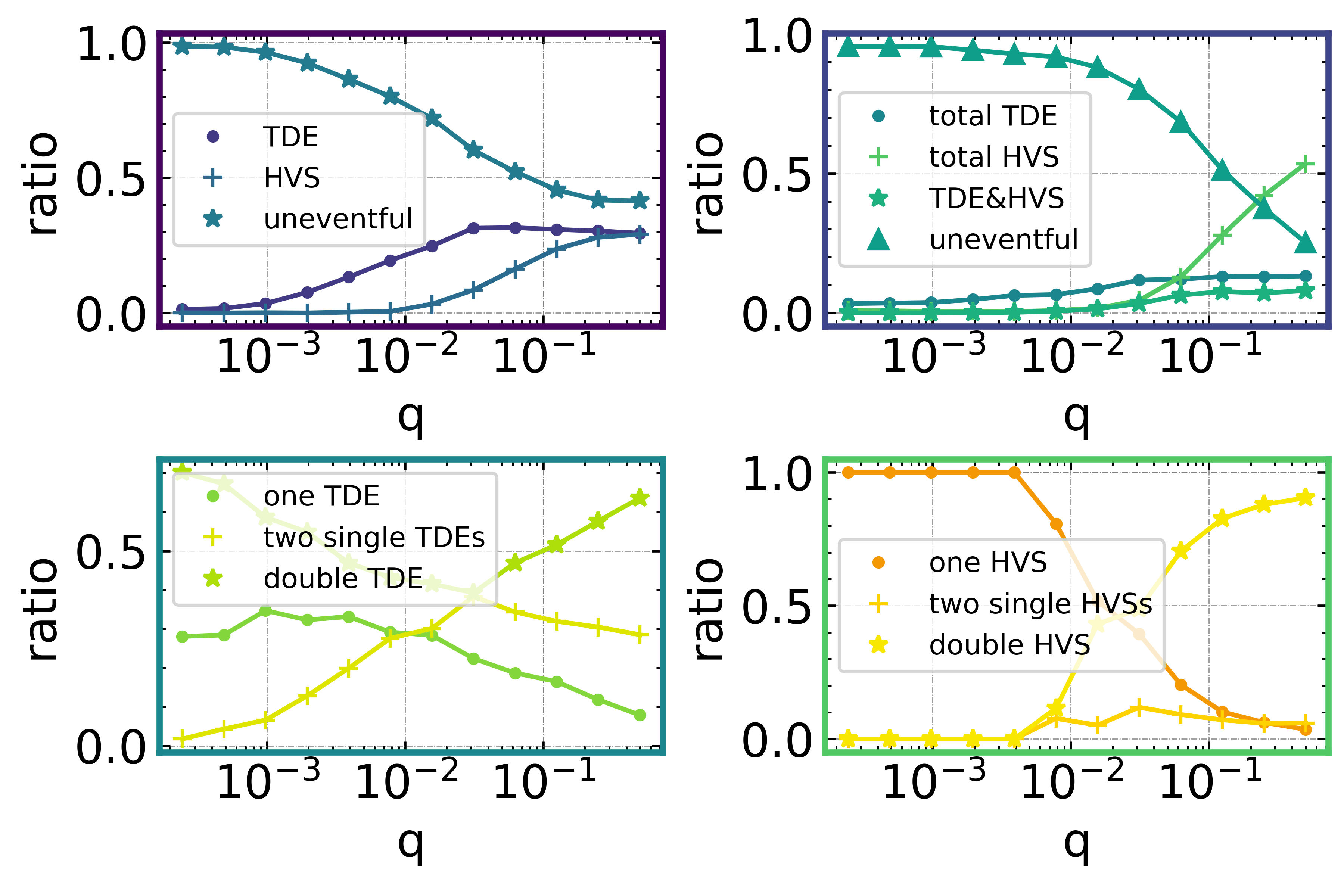} 
\caption{The relative rates of events as shown in
  Fig. \ref{fig:eventTree}. The top inset shows the rates for the
  upper branches of the tree in Fig.~\ref{fig:eventTree}. The middle
  panel shows the middle branches, and the bottom panel shows the
  lower branches. The color of each event box in
  Fig.~\ref{fig:eventTree} is identical to the corresponding event colour
  here. }
\label{fig:ratetree}
\end{figure}

Figure \ref{fig:ratetree} shows the relative rates for the different events shown in Fig. \ref{fig:eventTree} for $e_\bh = 0.5$. Both the TDE and HVS rates increase with increasing mass ratio $q$.

\begin{figure}
\centering
\includegraphics[width=\columnwidth]{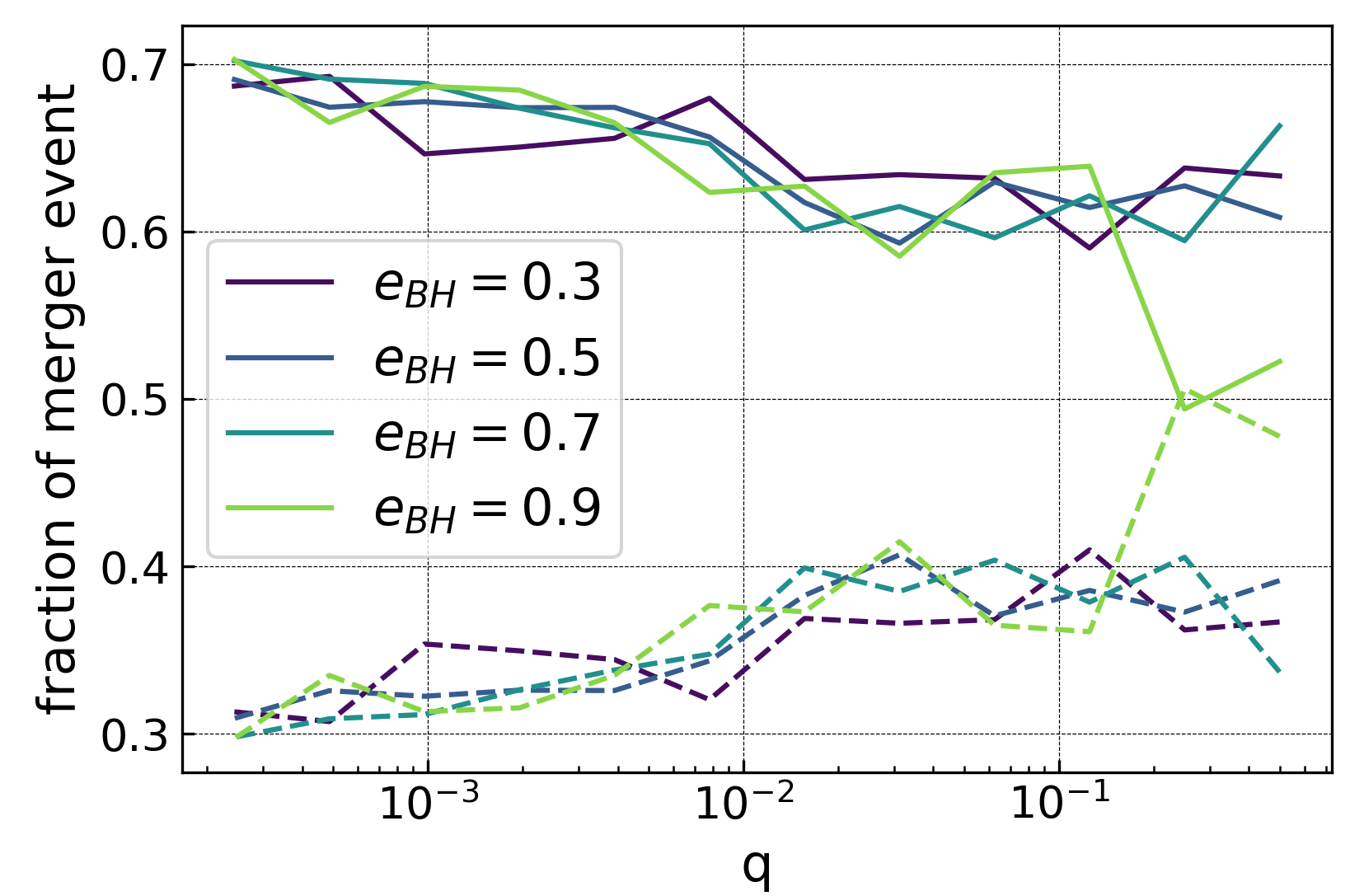} 
\caption{ The fraction of the event rate of stellar merging
  due to the secular effects versus non-secular or chaotic effects for different
  eccentricities of the SMBH binary. The solid lines show the fraction
  of the event rate of stellar merging driven by the inner LK oscillations. 
  The dashed lines show
  the fraction of the event rate of stellar merging due to the 
  non-secular or chaotic effects.}
\label{fig:mergerfraction}
\end{figure}

\subsubsection{Merger Rate as a function of Mass Ratio}
\begin{figure}
\centering
\includegraphics[width=\columnwidth]{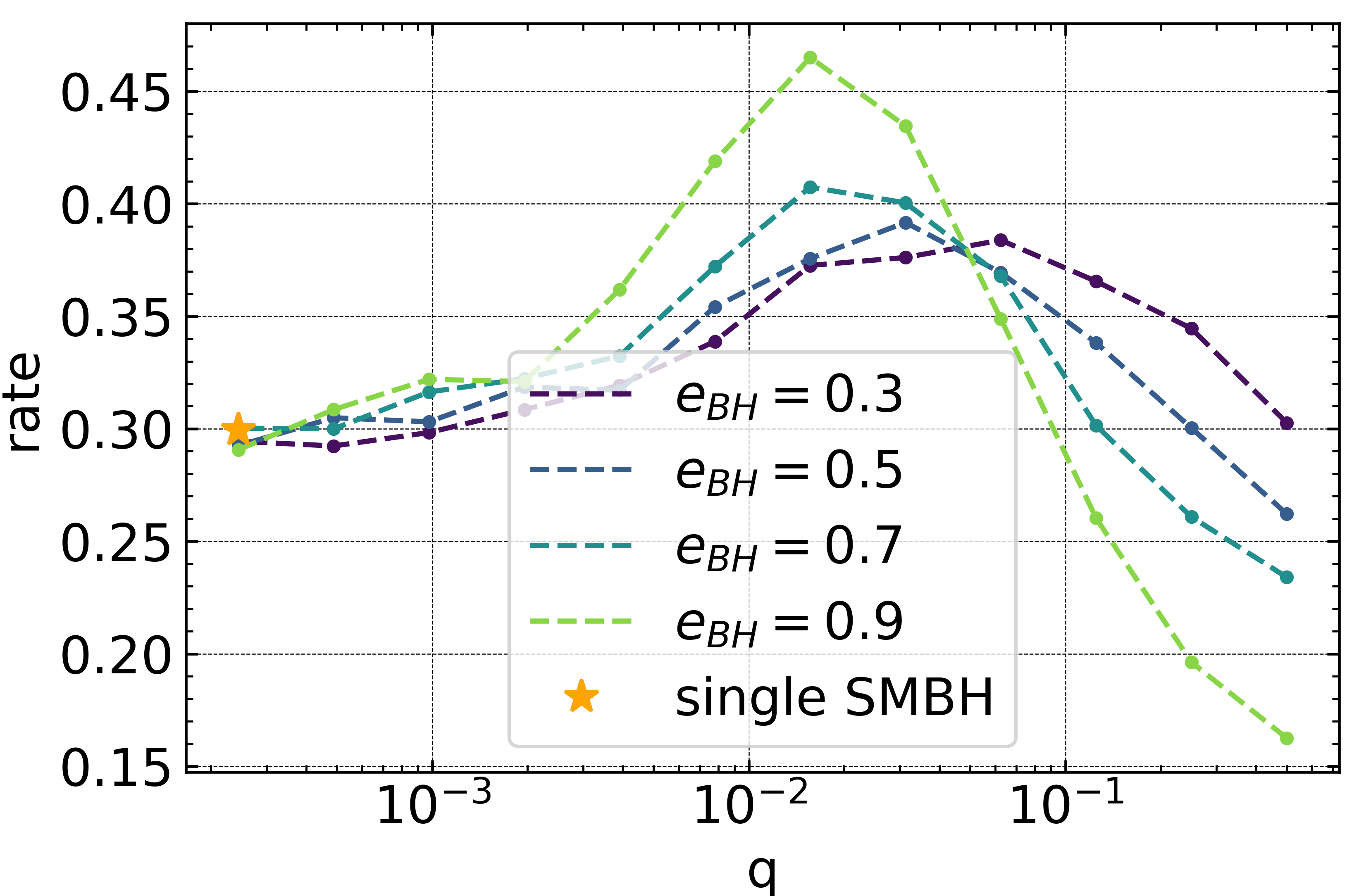} \\
\includegraphics[width=\columnwidth]{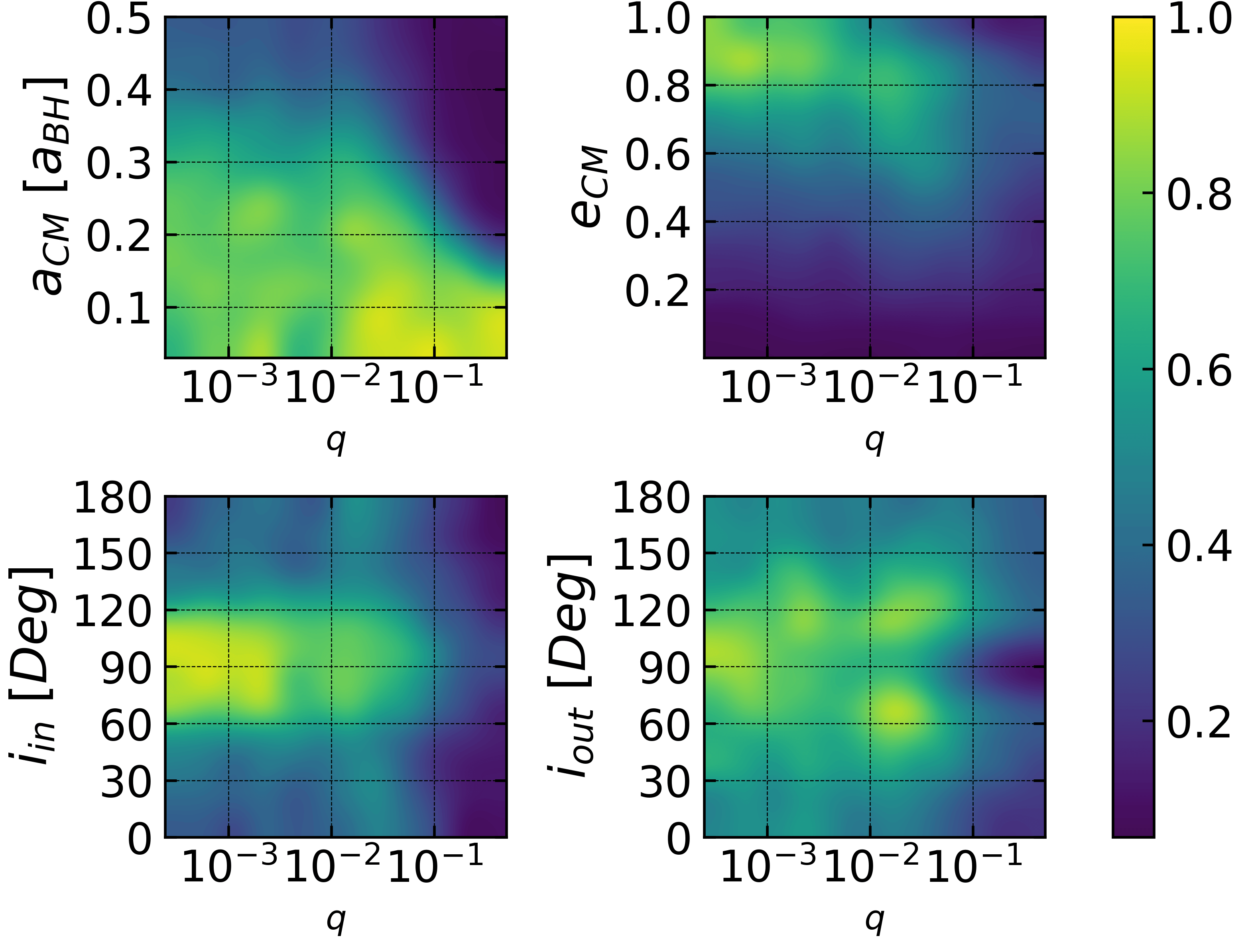}
\caption{Rates of mergers of stellar binaries. The upper panel shows
  the merger rates for different mass ratios $q$. The merger rate
  increases with increasing mass ratio $q$, at least until $m_2$
  becomes too massive and breaks the hierarchical structure of the
  four-body system. The colours in the bottom panel quantify the
  relative merger probability  in the indicated parameter
    space. }
\label{fig:stellarmerger}
\end{figure}

{ The most common event in our four-body system is stellar
  mergers. The merger events correspond to initial conditions for
  which the inner LK oscillations
  dominate. Fig.\ref{fig:mergerfraction} shows the fraction of merger
  events arising from different effects. We see that most merger
  events are due to LK oscillations. Therefore, the efficiency of the
  LK oscillations is what decides the dependence of the total merger
  rate on mass ratio $q$.} The upper panel of
Fig. \ref{fig:stellarmerger} shows the merger rates for different mass
ratio. This mechanism should be independent of the mass of the second
SMBH $m_2(q)$, since it is driven by the inner LK oscillations coming
from the primary SMBH. However, the upper panel of
Fig. \ref{fig:stellarmerger} indicates that the merger rate increases
slightly with increasing mass ratio. The reason for this slight
increase is that the inner and outer LK oscillations are not
completely independent. In regions where both the inner and outer LK
oscillations affect the stellar binary, the outer LK oscillations
could excite the stellar binary to migrate in closer to the primary
SMBH $m_1$, where the inner LK oscillations are stronger, leading to
higher merger rates.  The peak of the merger rate appears between $q
\sim 10^{-2}$ and $q \sim 10^{-1}$. Above the peak mass ratio, the
merger rate starts to decrease due to the four-body system more
frequently becoming unstable. Larger masses for the secondary SMBH
$m_2$ weaken both the inner and outer LK oscillations. Consequently,
in this case, the merger rate decreases with increasing mass ratio. At
the far left end, all lines converge to the star, which represents the
merger rate for the single SMBH case.

The lower panel of Fig. \ref{fig:stellarmerger} shows the relative probabilities for the occurrence of stellar mergers. The initial parameters considered in each of the four insets are $(a_\cm,q)$, $(e_\cm,q)$, $(i_{in},q)$ and $(i_{out},q)$. The colours quantify the relative probabilities in these parameter spaces. The first inset shows that mergers are more likely to occur at small $a_\cm$ and large $q$.  This is because, from Eq. \ref{eq:LKB} and Eq. \ref{eq:LKC}, smaller $a_\cm$ and larger $q$ translates in to stronger inner LK oscillations compared to the outer LK oscillations. The third inset shows that most mergers happen when the inclination $i_{in}$ nears $90^o$. This region is ideal for LK oscillations to operate effectively. The last inset confirms our earlier hypothesis, in the upper panel of Fig. \ref{fig:stellarmerger}, namely that the outer LK oscillations also contribute to mergers when $i_{out}$ even weakly concentrates at high inclination.

\subsubsection{HVS rate enhancement}\label{sec:hvsburst}
\begin{figure}
\centering
\includegraphics[width=\columnwidth]{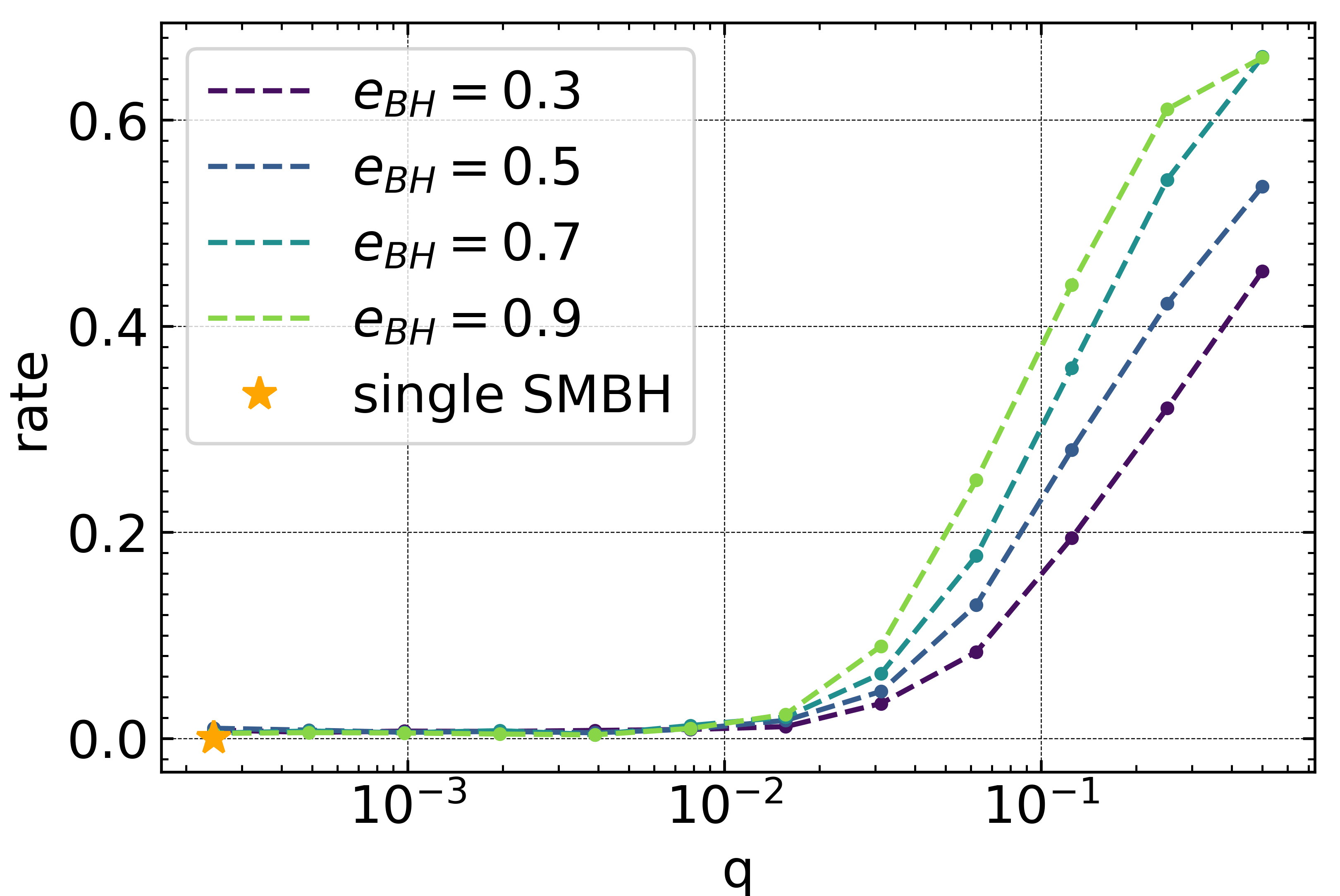} \\
\includegraphics[width=\columnwidth]{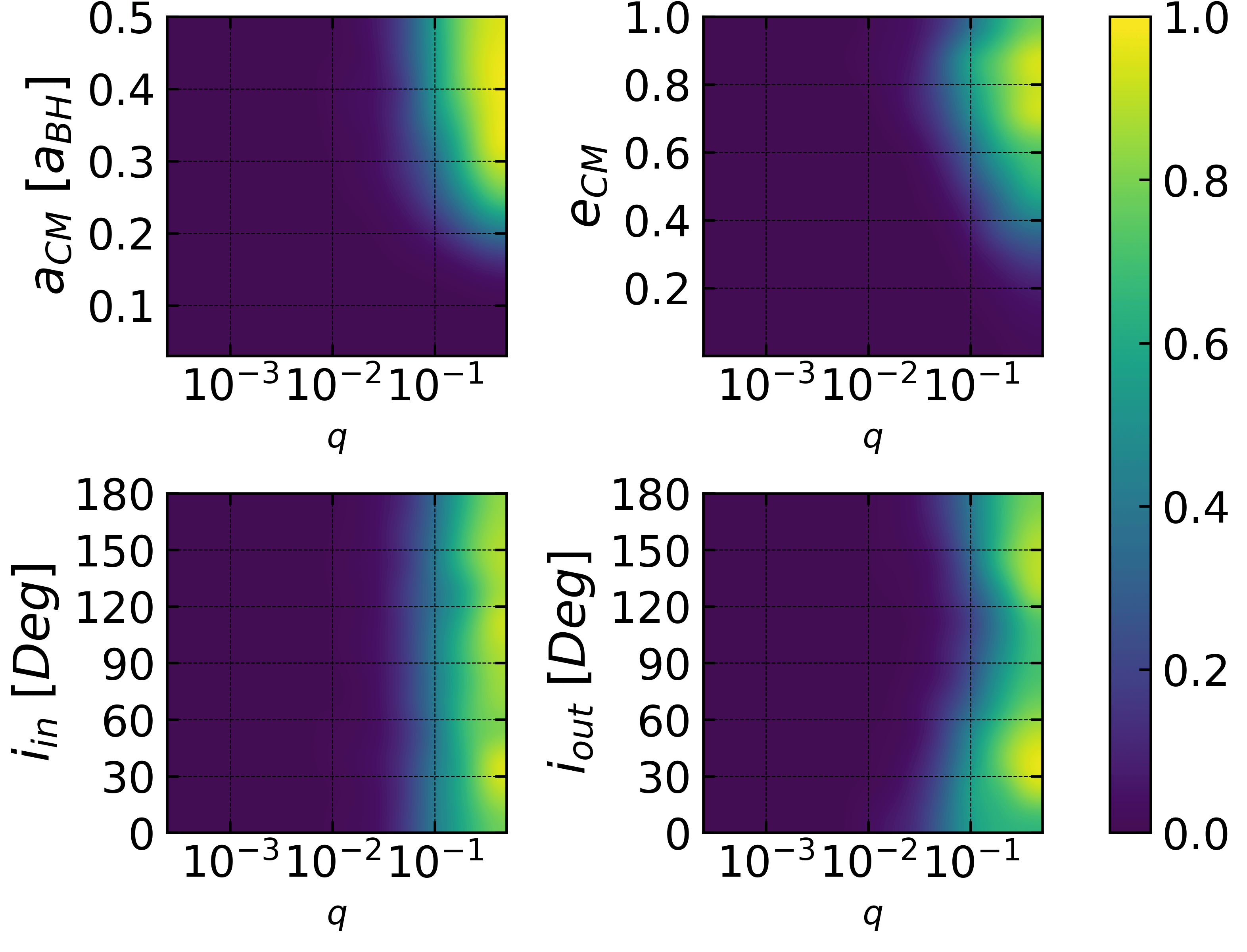}
\caption{Event rates of HVS versus the mass ratio of binary SMBH with different $e_\bh$. The upper panel shows that HVSs begin to be produced at $q\sim 10^{-2}$, and the rate increases at larger mass ratios. The lower panel shows the HVS rates in colour-space.  The yellow regions indicate the highest HVS rates and appear at large $q$, large $a_\cm$, large $e_\cm$ and near the critical angle for LK oscillations to occur, $i_c \sim 40^o or 140^o$.}
\label{fig:HVSburst}
\end{figure}

An interesting effect seen in our four-body system is a rapid increase
in the HVS rate with increasing mass ratio. The upper panel of
Fig. \ref{fig:HVSburst} shows the total HVS rates for different mass
ratios $q$ and eccentricities $e_\bh$. For very small mass
  ratios, the lines converge to the single SMBH case, and the HVS rate
  is nearly zero. Near the lowest mass ratios $q\sim 10^{-2}$, the
stability of the four-body system might be affected and breakup is
possible. At larger mass ratios, the HVS rate increases significantly
with increasing mass ratio $q$. This can potentially be explained by
the strong perturbation induced from the secondary SMBH $m_2$ on the
stellar binary, which could transfer energy and angular momentum to
the binary.  This could even result in the collisional ejection of the
binary. The bottom right panel of Fig. \ref{fig:HVS} shows a good
example of this process. As is clear, a larger secondary SMBH mass
$m_2$ along with a closer distance from the primary SMBH (i.e., larger
$e_\bh$) result in stronger perturbations, which translates in to an
elevated rate of HVS production.

The lower panel of Fig. \ref{fig:HVSburst} shows that HVSs are more
likely to be produced at larger $a_\cm$. This is because binaries at
smaller $a_\cm$ are more likely to be consumed as TDEs due to LK
oscillations. This panel also indicates that the HVS rate is
independent of the outer inclination $i_{out}$, but increases near the
critical angle for LK oscillations, namely $i_c \sim 40^o$ and $140^o$
for $i_{out}$.

\begin{figure}
\centering
\includegraphics[width=\columnwidth]{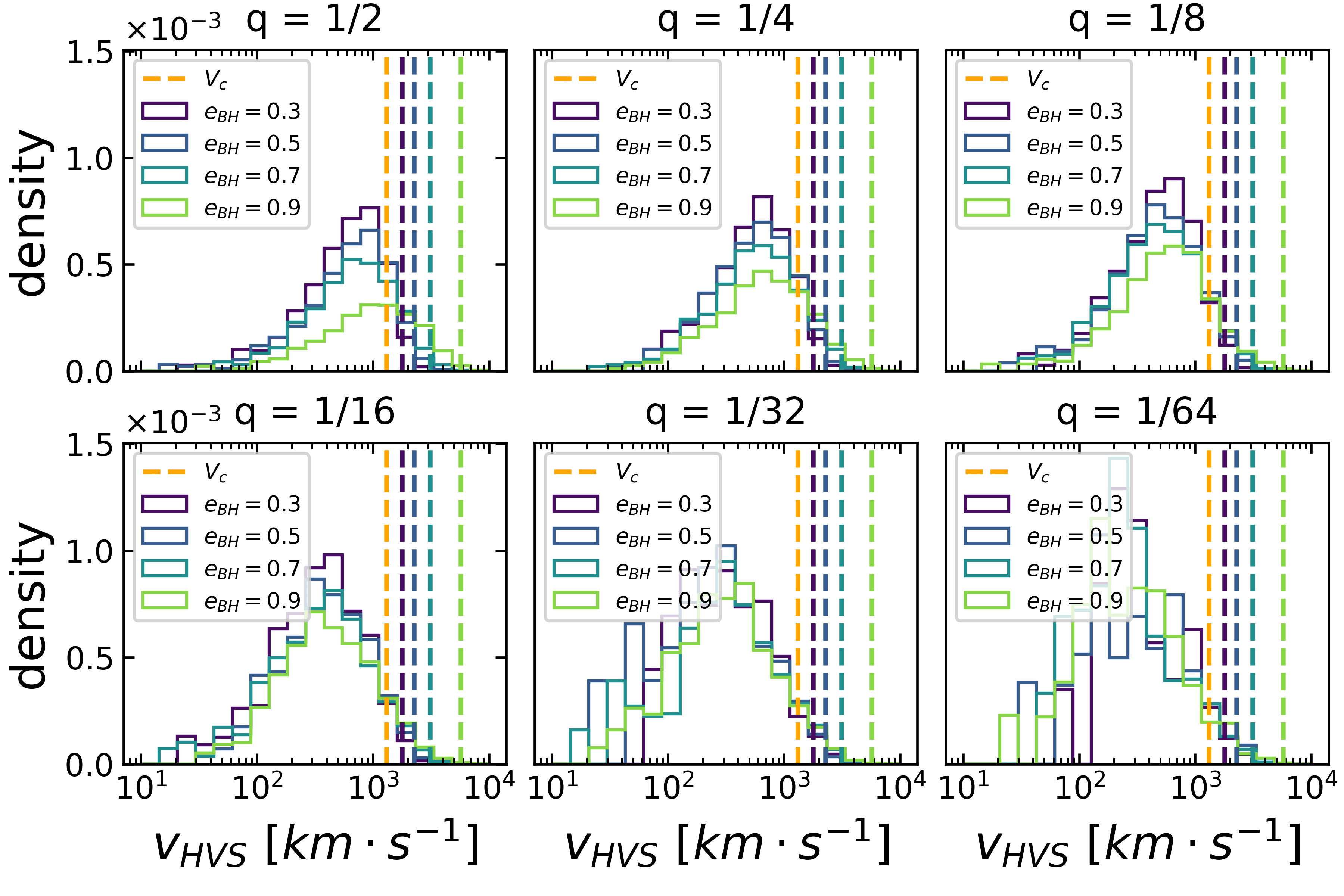}
\caption{Distribution of the normalized HVS velocity at escape radius
  $60\,a_\bh$ from the centre of mass of the SMBH binary, for
  different mass ratios $q$ and eccentricities $e_{\rm BH}$ of the
  binary SMBH.  Here $V_c = \sqrt{Gm_1/a_\bh}$ is the binary orbital
  velocity of the SMBH binary. The dashed lines show, for each binary
  model with a given eccentricity, the velocity of the secondary SMBH
  at the pericenter.}
\label{fig:HVSdist}
\end{figure}

Figure \ref{fig:HVSdist} shows the HVS velocity distribution 
measured at the escape distance, or $60\,a_\bh$ from the centre of mass of the
SMBH binary. The typical escape velocity is $\sim 300-1000$~km~s$^{-1}$, which
fits the observations very well.

\begin{figure}
\centering
\includegraphics[width=\columnwidth]{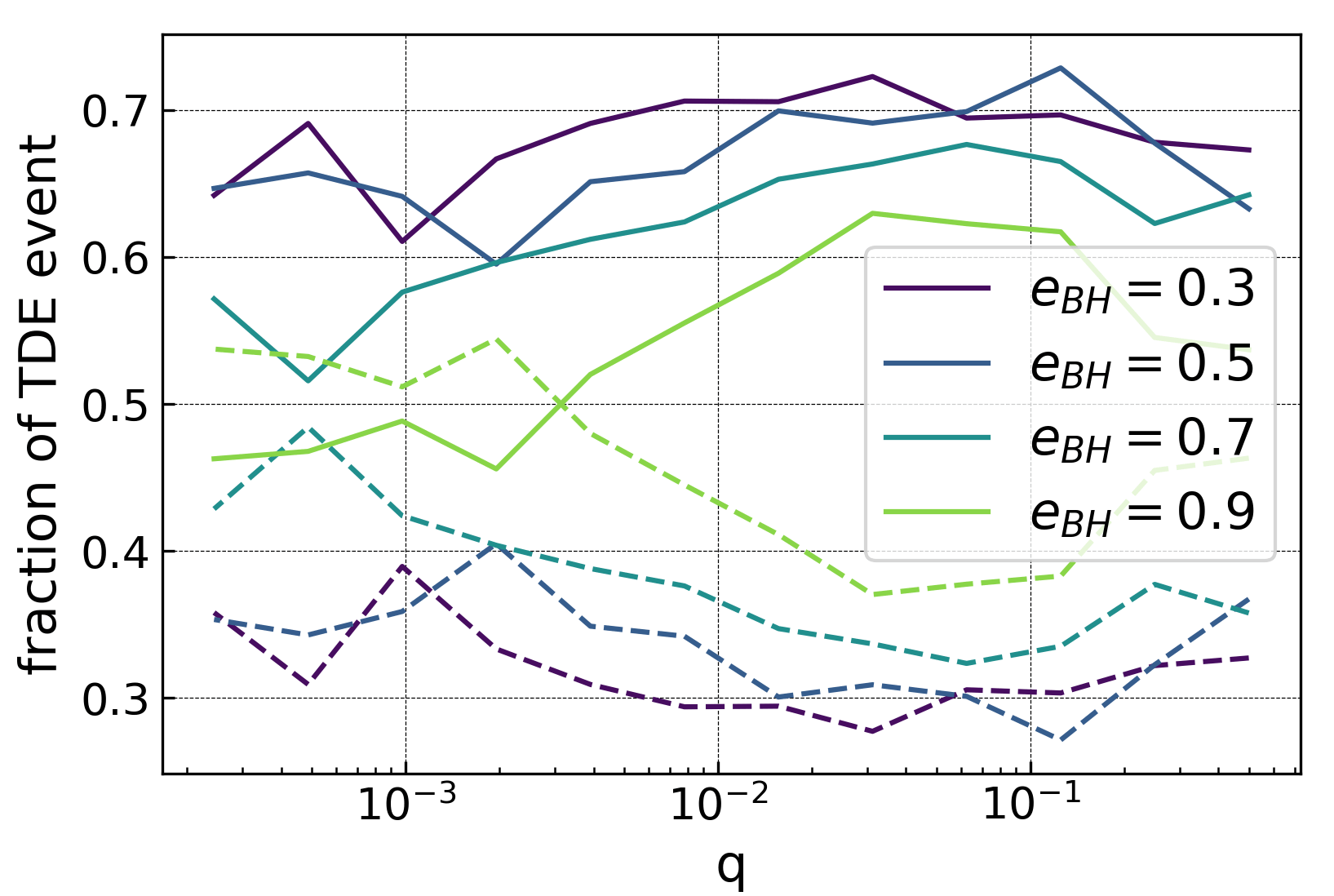} 
\caption{ The fraction of TDEs due to the secular
  effects versus the non-secular or chaotic effects as a function of the
  mass ratio of the SMBH binary for different eccentricities of the SMBH binary. 
  The solid lines show the fraction
  of TDEs driven by the outer LK oscillations. The dashed lines show the
  fraction of TDEs due to the non-secular or chaotic effects.}
\label{fig:TDEorigin}
\end{figure}

\subsubsection{TDE rate enhancement}
\begin{figure}
\centering
\includegraphics[width=\columnwidth]{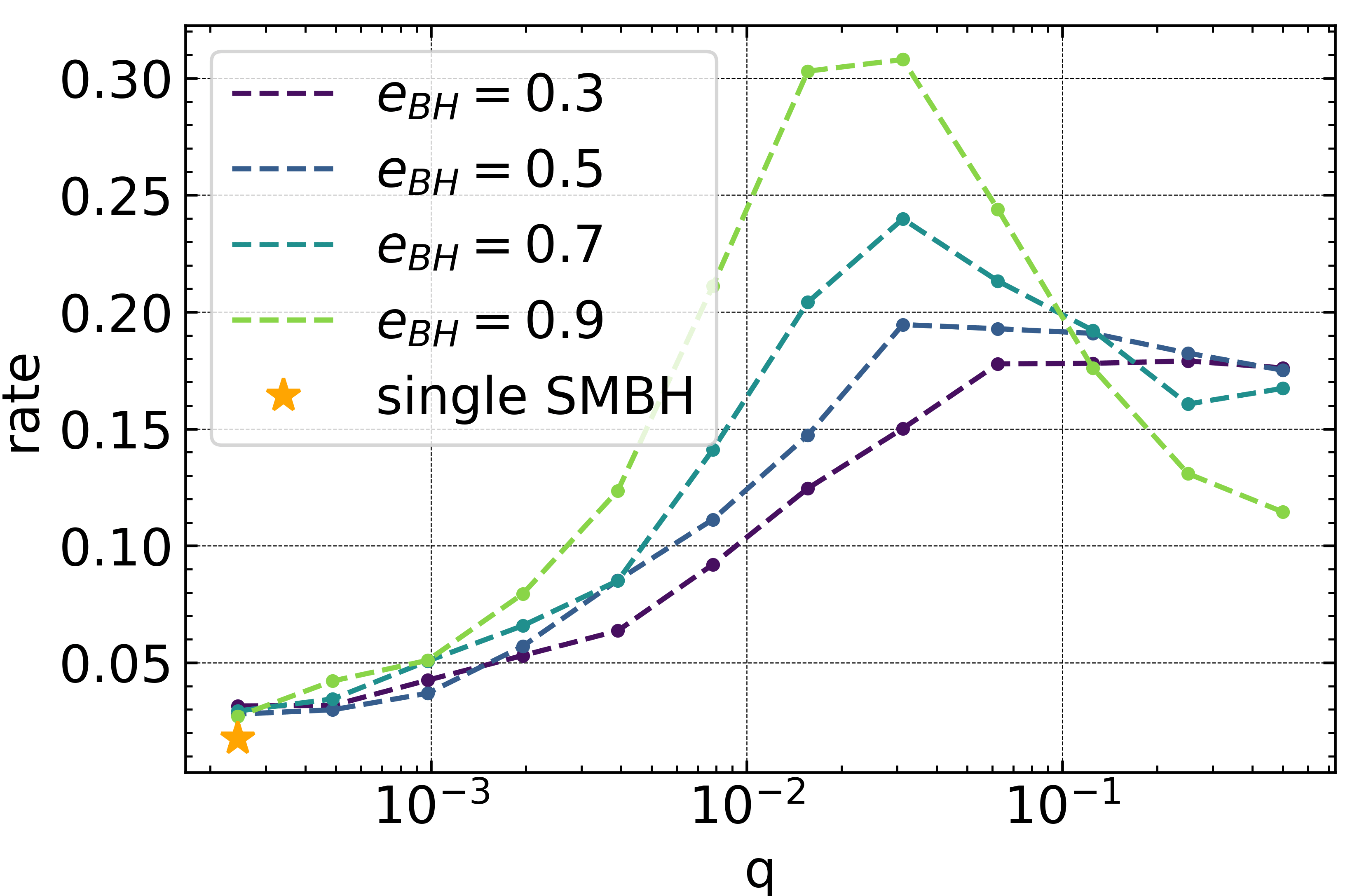} \\
\includegraphics[width=\columnwidth]{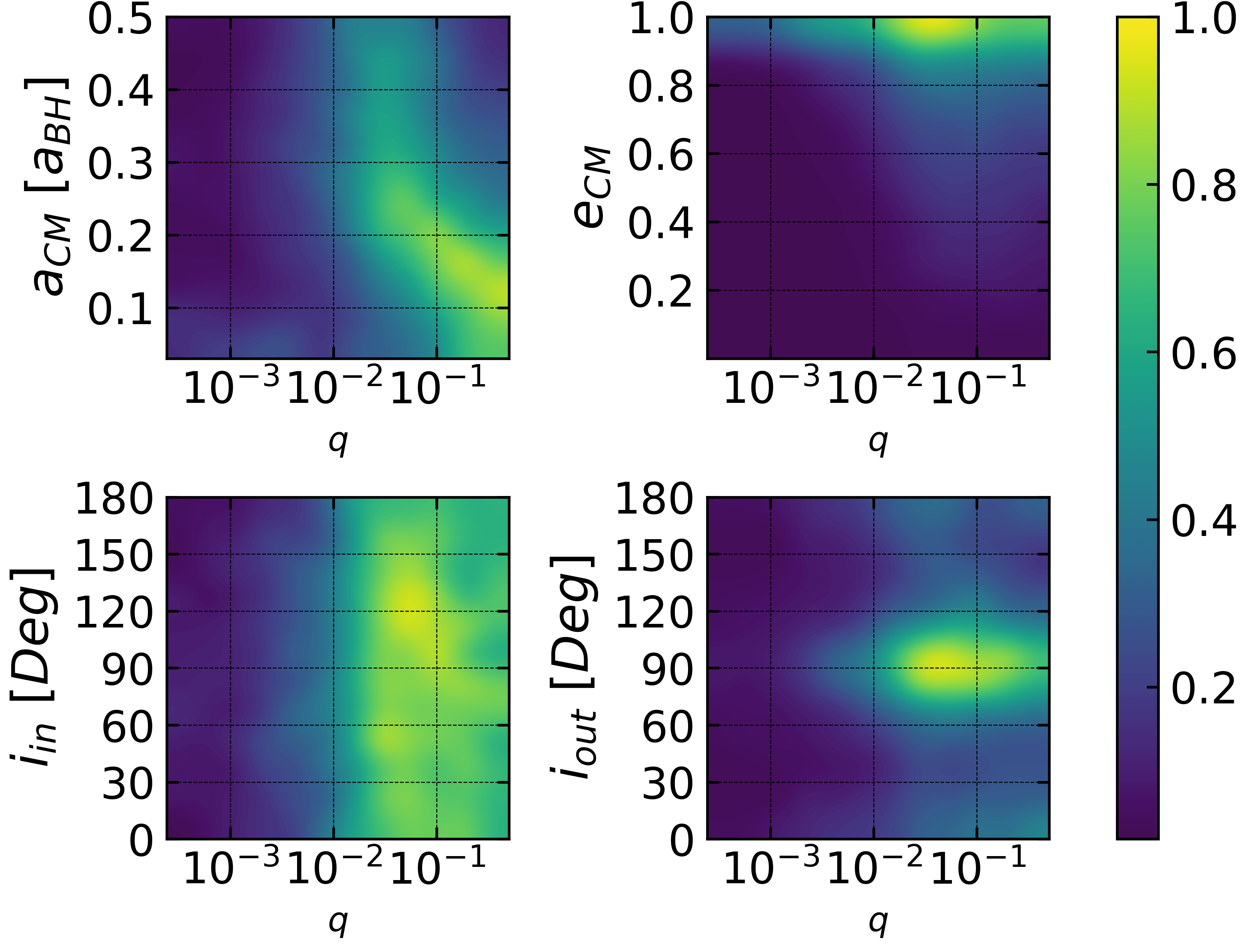}
\caption{TDE rates versus mass ratio of the binary SMBH with different $e_\bh$. The upper panel illustrates a significant enhancement in the TDE rates at larger mass ratios $q$, at least up to $\sim 0.03$. Chaotic effects are responsible for the TDE rate decrease a larger values of $q$.  The lower panel illustrates the primary effect we have attributed to the outer LK oscillations - TDE rates are high at high inclination $i_{out}$ for the outer orbit, and independent of the inner inclination $i_{in}$. The first inset, which plots $a_\cm-q$, shows the stability boundary for hierarchical triples. }
\label{fig:TDEenhancement}
\end{figure}

{ Fig.\ref{fig:TDEorigin} shows the fraction of TDEs driven
  by secular versus non-secular effects. We see that the relative
  fractions are sensitive to the eccentricity of the SMBH binary.
  With increasing $e_\bh$, non-secular and chaotic effects become
  significant. The relative fraction has only a weak dependence on the
  mass ratio of the SMBH binary. However, the total TDE rate is
  enhanced significantly due to the presence of the secondary SMBH.
  LK oscillations driven by the outer orbit are the main source of the
  increase in the TDE rate.}  Equation \ref{eq:LKC} shows the outer LK
oscillation time-scale$\sim (1-e^2_\bh)^{3/2}/q$.  Both the mass ratio
$q$ and the eccentricity $e_\bh$ increase quickly, along with a
decrease in the time-scale for LK oscillations, leading to enhanced
orbital excitation in the outer triple. This explains the observed
increase in the TDE rate for $q\lesssim 10^{-2}$ (see the upper panel
of Fig. \ref{fig:TDEenhancement}). As the mass ratio $q$ increases,
the stellar binary begins to feel the effects of the increasingly
massive secondary SMBH $m_2$. Consequently, the LK oscillations in the
outer triple are gradually reduced by the presence of the secondary
SMBH $m_2$. This is the reason for finding that the TDE rate does not
increase and can even decrease at larger mass ratios. The TDE rate for
the single SMBH case at the far left end drops from $\sim 3\%$ to
$\sim 1.5\%$. This is due to the gap in the single TDE rate from
Fig. \ref{fig:allrates}.

The second SMBH $m_2$ enhances the TDE rate significantly, from $2.8\%$ to $17.6\%$ for $e_\bh=0.3$ and from $2.8\%$ to $30.8\%$ for $e_\bh =0.9$. The LK oscillations require that the outer triple orbit must have an outer inclination $40^\circ < i_{out} <  140^\circ$. The lower panel of Fig. \ref{fig:TDEenhancement} shows these effects mediated by LK oscillations. Most TDEs appear at high inclinations $i_{out}$, when the outer LK oscillations dominate. This panel also clearly illustrates that the TDE rate is independent of the inner inclination $i_{in}$.

\subsubsection{HVS with TDE}
A special event in this four-body system is the simultaneous occurrence of an HVS with a TDE. The stellar binary is disrupted when $r_*$ reaches the Hill radius $r_{Hill}$. One of the stars then gets ejected as an HVS, while the other star falls into the tidal radius $r_{t*}$ of one of the SMBHs (usually the primary, but see Figure \ref{fig:TDEfromSecond}). The binary disruption plays a fundamental role in this special event.  As discussed in Section \ref{sec:hvsburst}, in certain regions of the relevant parameter space, the secondary SMBH can transfer energy and angular momentum to the stellar binary.  Subsequently, the binary can be collisionally ejected from the system.  If the distance of the stellar binary to the primary SMBH $m_1$ remains sufficiently large, the stellar binary can remain bound and be ejected as a bound object.

\begin{figure}
\centering
\includegraphics[width=\columnwidth]{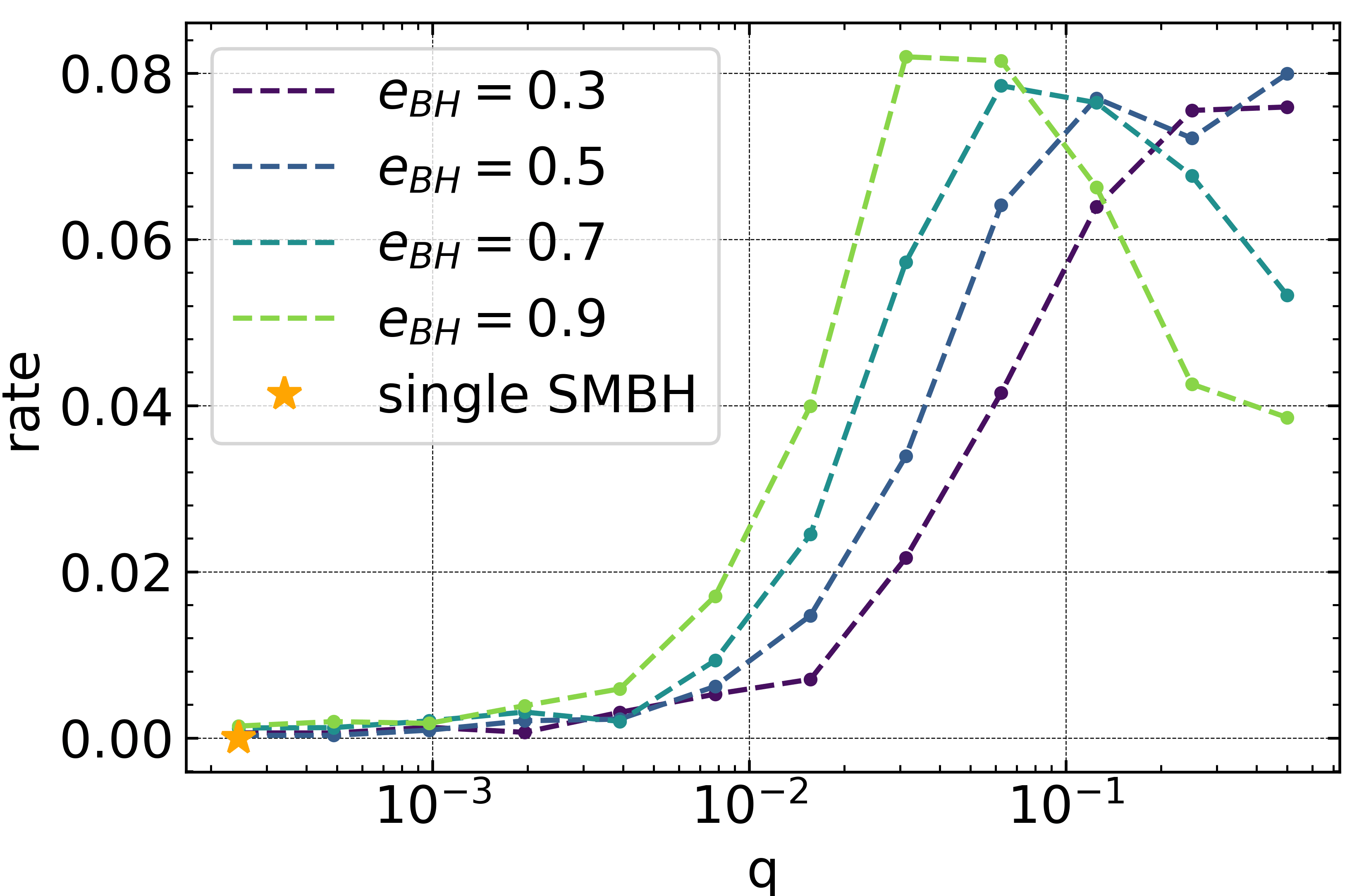} \\
\includegraphics[width=\columnwidth]{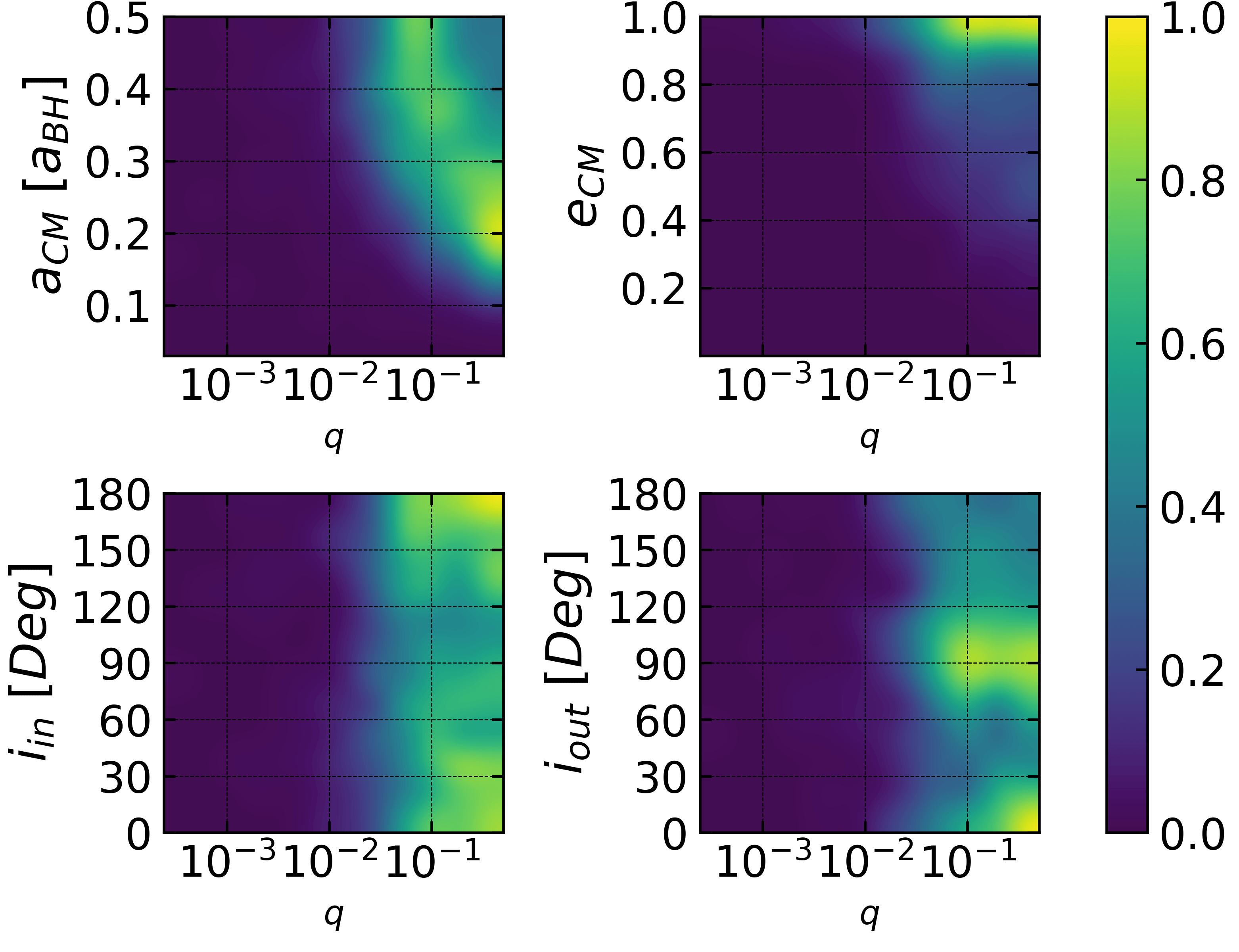} 
\caption{The rates for the simultaneous production of an HVS and a TDE
  as a function of the mass ratio of the binary SMBH with different
  $e_\bh$. The upper panel shows the presence, albeti wit low probability, of this special event in
  our four-body system. The bottom panel shows the relative
  probability for the occurrence of this event in each parameter
  space. The TDE\&HVS combination tends to occur at high inclination
  angles for the outer orbit $i_{out}$ and low inclination angles for
  the inner orbit $i_{in}$.}
\label{fig:HVSwithTDE}
\end{figure}

Figure \ref{fig:HVSwithTDE} shows the event rate for the simultaneous formation of an HVS and a TDE (i.e., HVS\&TDE combination), for different eccentricities $e_\cm$. The simulated data resemble what we would naively expect from a simple superposition of the TDE and HVS data, albeit with a few additional modifications. 
Although the dependence of energy and angular momentum redistribution within the four-body system on the mass ratio $q$ is not completely understood, we can still draw reliable conclusions from the lower panel of Fig. \ref{fig:HVSwithTDE}. In mass ratio-space, the HVS\&TDE combination looks very similar to what is seen for the TDE case alone, with the exception that the former outcome extends out beyond this region to the upper right. The inner inclination $i_{in}$ concentrates at low inclination angles, such that no binary merger events occur. The outer inclination $i_{out}$ concentrates at high inclination angles, such that the decoupled star can be excited in to the tidal disruption radius.

\subsection{TDE from the Secondary SMBH}
Both the Schwarzschild radius $R_s$ and the tidal disruption radius for stars $r_{*t}$ increase with increasing SMBH mass. The tidal disruption radius $r_{*t}$ is given by Eq. \ref{eq:TDE}, and the Schwarzschild radius is
\begin{equation}
R_s = \frac{2Gm_\bh}{c^2}\,.
\end{equation}
Above a critical BH mass $m_{\rm BH,crit}$, the Schwarzschild radius
becomes greater than the tidal disruption radius. Above this critical
SMBH mass it should not be possible to observe any TDEs, because the
SMBH will swallow the whole stars and no light will escape from within
$R_s$.
In the limit of a primary SMBH with mass $m_1 > m_{\rm BH,crit}$, but
a secondary with mass $m_2 < m_{\rm BH,crit}$, it is only the
secondary SMBH that can cause TDE events.  By incorporating the
empirical $M_{\bh}-M_{\rm gal}$ relation
\citep[.e.g][]{Ho,georgiev16}, this opens up the possibility to use
TDEs as detectors for secondary SMBHs/IMBHs orbiting in the nucleus of
a far away galaxy.  This is because, since the $M_{\bh}-M_{\rm g}$
relation increases monotonically, it can be used to convert the
critical BH mass $m_{\rm BH,crit}$ in to a critical galaxy mass
$M_{\rm g,crit}$, above which no TDEs should be produced by the
central SMBH.  However, this assumes that the central massive BH in
every galaxy is isolated.  Cosmological simulations, on the other
hand, suggest that both major and minor mergers between galaxy pairs
occur frequently, and in the process deliver their own central SMBHs
to the nuclear regions of the product of this galaxy-galaxy merger.
This could predict the presence of a large population of secondary
SMBHs in galaxies, in presumably stable orbits about the primary SMBH.

\begin{figure}
\centering
\includegraphics[width=\columnwidth]{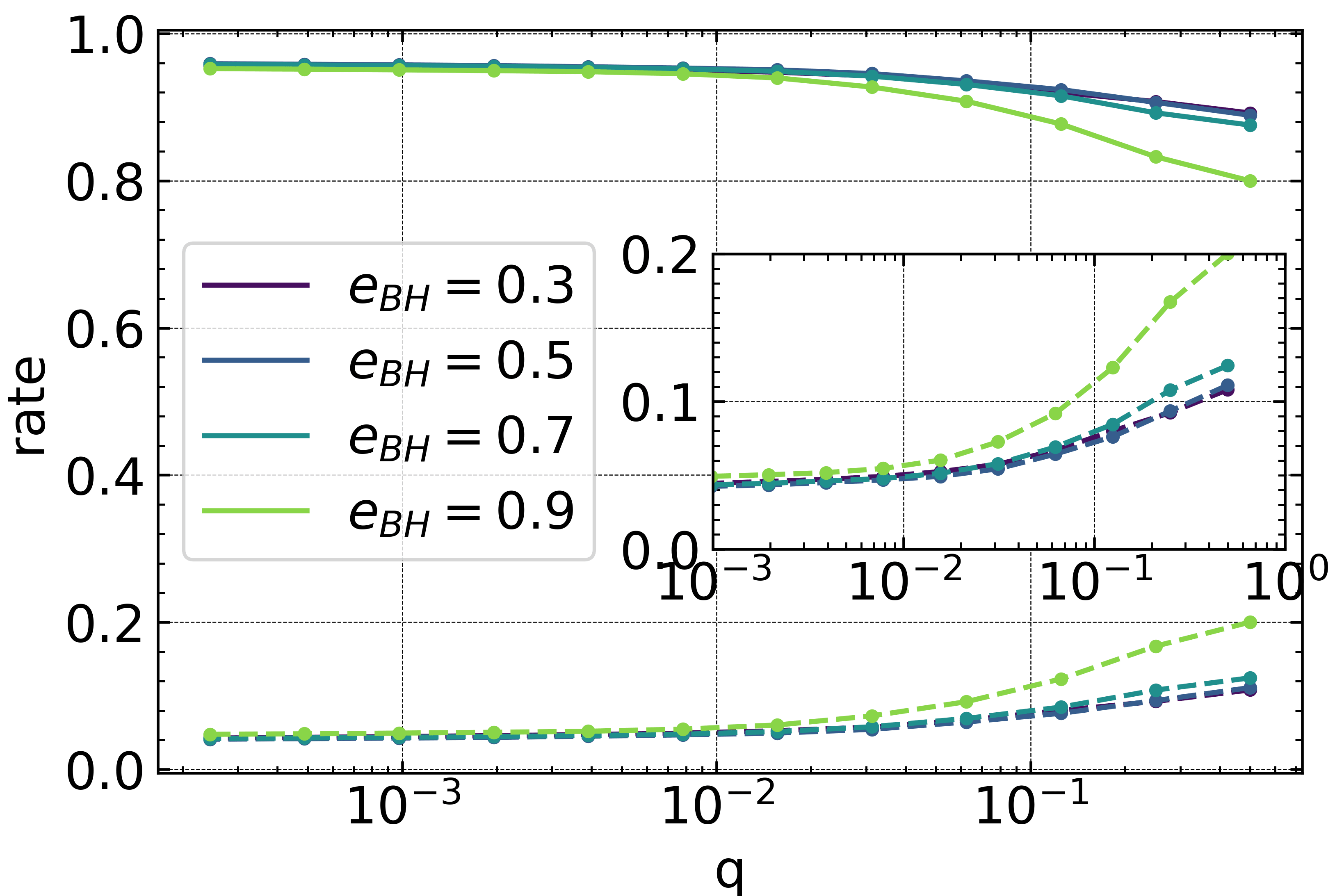} 
\caption{The event rates of TDE by both the primary and the secondary
  black holes as function of the mass ratio of binary SMBH.  The solid
  lines show the fraction of TDEs occurring around the primary SMBH.
  The dashed lines show the fraction of TDEs occurring around the
  secondary SMBH. The curves at different eccentricity all converge to
  a few percent below $q\sim 10^{-3}$, such that the ratio becomes
  roughly independent of $q$.  At large mass ratios (i.e., after
  $q\sim 10^{-2}$), the fraction of TDEs from the secondary SMBH
  quickly increases. }
\label{fig:TDEfromSecond}
\end{figure}

The presence of the secondary SMBH in orbit around the primary SMBH
opens the door to observing TDEs from the secondary SMBH.  But we do
not know how often this should occur, if ever.  To answer this
question, Figure \ref{fig:TDEfromSecond} shows the relative
probability for a secondary-induced TDE for different mass ratios $q$
and different eccentricities $e_\bh$.  As is clear, the secondary
always contributes to the total rate of TDEs, but only by a few
percent at very low mass ratios.  At larger mass ratios, the fraction
of TDEs produced by the secondary increases steadily, converging
toward 50\% at $q = 1$.  Importantly, independent of the mass ratio,
the secondary always contributes a base minimum of a few percent to
the total TDE rate.  Extrapolating this result to even lower mass
ratios suggests that even stellar-mass BHs closely orbiting a central
SMBH have a non-negligible probability of causing a TDE.  This begs
the question: If $\sim 100$ stellar-mass BHs are in close orbits about
a central massive SMBH, and each one should contribute of order a
percent to the total observed TDE rate, then does the TDE rate from
the SMBH become nearly equal to the TDE rate from the swarm of
stellar-mass BHs?  We intend to explore this interesting question in
more detail in future work.

\section{Summary} 
\label{discussion}
In this work, we have studied the fate of main sequence stellar binaries
orbiting around a primary central SMBH perturbed by a remote secondary
SMBH, as a function of the mass ratio of SMBH-SMBH binaries. The presence
of the secondary SMBH significantly changes the evolution of
these stellar binaries, and the relative rates of observable
astrophysical phenomena. We have performed $N$-body simulations of this
four-body system with different SMBH-SMBH binary mass ratios, and
analyzed the fate of the system. Our main conclusions can be summarized
as follows.

Our simulations show that the total event (TDE, HVS, merger) rates are
 increased by the presence of a secondary SMBH $m_2$, and continue 
 to increase as $m_2$ increases (Figure~\ref{fig:allrates} indicates that 
 the fraction of uneventful simulations continuously decreases as the mass ratio increases). 
 The presence of the secondary SMBH acts to migrate the stars toward the primary 
 SMBH via the outer LK oscillations, where they are consumed by TDEs, 
 HVSs and mergers. This suggests that galaxies observed to exhibit higher rates of extreme 
 astrophysical events (i.e., TDEs, HVSs, mergers) are more likely to harbour an SMBH binary.

Merger events mainly occur due to the inner LK oscillations. Equation~\ref{eq:LKB} 
shows the timescale for the inner LK oscillations to operate. For a given
 stellar mass density profile (i.e., given a distribution in $a_\cm$), the merger rate 
 in a galaxy is determined by the properties of its primary SMBH. However, our simulations 
 show that the secondary SMBH could slightly increase the merger rate in 
 the range $q \sim < 10^{-2}$. In this mass ratio range, the secondary SMBH transports 
 more stars to the inner regions near the primary SMBH.  Here, the inner LK 
 oscillations are stronger, leading to higher merger rates.

HVSs are produced from the decoupling of the stellar binary at its
break up radius $r_{\mathrm{bt}}$ due to the primary SMBH. Previous
work indicates that this decoupling tends to produce HVSs by ejecting
one of the stars in the binary. Our simulations show that the rate for
this process to operate is very low (see the rates of single HVSs in
Figure~\ref{fig:allrates}). In this work, we have identified a more
efficient way to produce HVSs, namely via strong perturbations from
the secondary SMBH.  These strong perturbations can even eject the
stellar binary at pericenter without unbinding the binary. The rate
for this process increases rapidly above mass ratios $q\ge 10^{-2}$
(see Figure~\ref{fig:HVSburst}). These `double HVSs' should only be
produced by SMBH binaries, with a rate much higher than for single
HVSs.\footnote{Note that this does not account for direct interactions
  between a single SMBH and stellar triple systems
  \citep[e.g.][]{perets12}, but the rate of these interactions should
  be very low \citep{leigh13}.} Hence, hypervelocity binary star
systems could be regarded as a smoking gun for the presence of an SMBH
binary\citep{2007ApJ...666L..89L,Sesana09}. Incidentally, one such system has
possibly been identified in the Milky Way \citep{brown10}.
Figure~\ref{fig:HVSdist} shows the HVS velocity distribution evaluated
at a distance of $60~a_\bh$ from the SMBH primary. The typical
velocities range from $300~{\rm km}~{\rm s}^{-1}$ to $1000~{\rm
  km}~{\rm s}^{-1}$, depending on the mass ratio of the SMBH-SMBH
binary. { To compare the distribution of HVS velocities with
  observations more precisely, the Galactic potential must be taken
  into consideration, which will be done in a follow up paper (Wang et
  al. in prep.).  In turn, the Galactic potential can be constrained
  from the observed distribution of HVSs using the upcoming data from
  the Gaia satellite, as discussed in
  \citep{Kenyon2014,Fragione2017,Marchetti2017}.}

The TDE rate is quite low for the single SMBH case, relative to the
SMBH-SMBH binary case.  This is because the secondary SMBH acts to
migrate stars to the inner region around the primary SMBH via strong outer
LK oscillations. This effect could focus the stars to the centre of
the galaxy, close to the primary SMBH, and in the process accelerate
the rate of tidal disruption events. The secondary SMBH increases the
TDE rate significantly at large mass ratios $q$. Therefore, this opens
up the possibility of using the relative rates of TDEs and HVSs to
observationally constrain the occurrence of SMBH-SMBH binaries in
galactic nuclei.

TDEs often occur due to disruption by the secondary SMBH, instead of
the primary.  This opens up the possibility of observing TDEs in
massive galaxies with the most massive central SMBHs, when no TDEs
should be expected.  This is because, via the M-$\sigma$ relation, the
host SMBH should have a mass above the critical mass for TDEs to occur
inside the Schwarzschild radius $R_{\rm S} > r_{*,t}$.  Hence, no TDEs
should be produced by the most massive SMBHs in the most massive
galaxies.  Thus, the observation of even a single TDE in such a
massive galaxy would be the smoking gun of a secondary lower mass SMBH
companion.

Our results show that the presence of a secondary SMBH changes the
relative rates of TDEs and HVSs relative to the isolated SMBH case.
Hence, observations of these two phenomena could help to constrain the
possible presence of a central IMBH in the Galactic Centre, and/or the
presence of SMBH-SMBH binaries in the nuclei of other
galaxies. The dependence of the event rates in our
  simulations on the SMBH-SMBH binary mass ratio could potentially be
  used to constrain the frequency of SMBH-SMBH binaries in galactic
  nuclei, and even their mass ratio distribution.  To this aim, a
follow up paper (Wang et al. in prep.)  will be devoted to a detailed
comparison between simulations and all the available observational
data.

\section*{Acknowledgments}
YFY is supported by National Natural Science Foundation of China (Grant No. U1431228, 11725312, 11233003, and 11421303).
Results in this paper were obtained using the high-performance LIred computing system at the Institute for Advanced 
Computational Science at Stony Brook University, which was obtained through the Empire State Development grant NYS \#28451.

\label{lastpage}

\newpage
\clearpage
\appendix
\section{\\Symplectic Algorithm}
The advantage of a symplectic algorithm lies in its ability to conserve the Hamiltonian structure of the system. Non-symplectic algorithms control the error to remain within a specified tolerance level, by reducing the time step length. However, if the integration time is long enough, the systematic error will accumulate non-negligibly. The rate of accumulation depends on the adopted time step length. For most gravity integrators, it is adequate to simply adjust the time step length to an acceptable value in order to ensure that the rate of error accumulation is negligible. This is because very long integration times are typically not required. However, to study systems with multi-scale dynamics (e.g., in our four-body system, the spatial scale of the SMBH binary is $\sim \pc$, but the spatial scale of the stellar binary is $\sim \au$), we need to make sure that we are not failing to resolve important physics, and correctly allowing for the various subtle dynamical processes characteristic of many-body systems to exert their impact on the evolution of the system. This usually requires very long integration times (e.g., the time scale for effect A to operate is much too long for effect B to have any effect, such that the shortest time-scale is what ultimately decides the integration time). 

In systems with characteristic multi-scale dynamics, non-symplectic algorithms need extremely long integration times to accurately resolve the relevant physics at the minimum scale. Due to its symmetric numerical format, symplectic algorithms conserve the Hamiltonian structure for much longer integration times. Therefore, over long integration times, symplectic algorithms perform better than non-symplectic algorithms.  Unfortunately, the precise symmetry applied to the numerical formatting depends on the adopted time step. This means that only algorithms that use a fixed step length are truly symplectic. 

\begin{figure}
\centering
\includegraphics[width=\columnwidth]{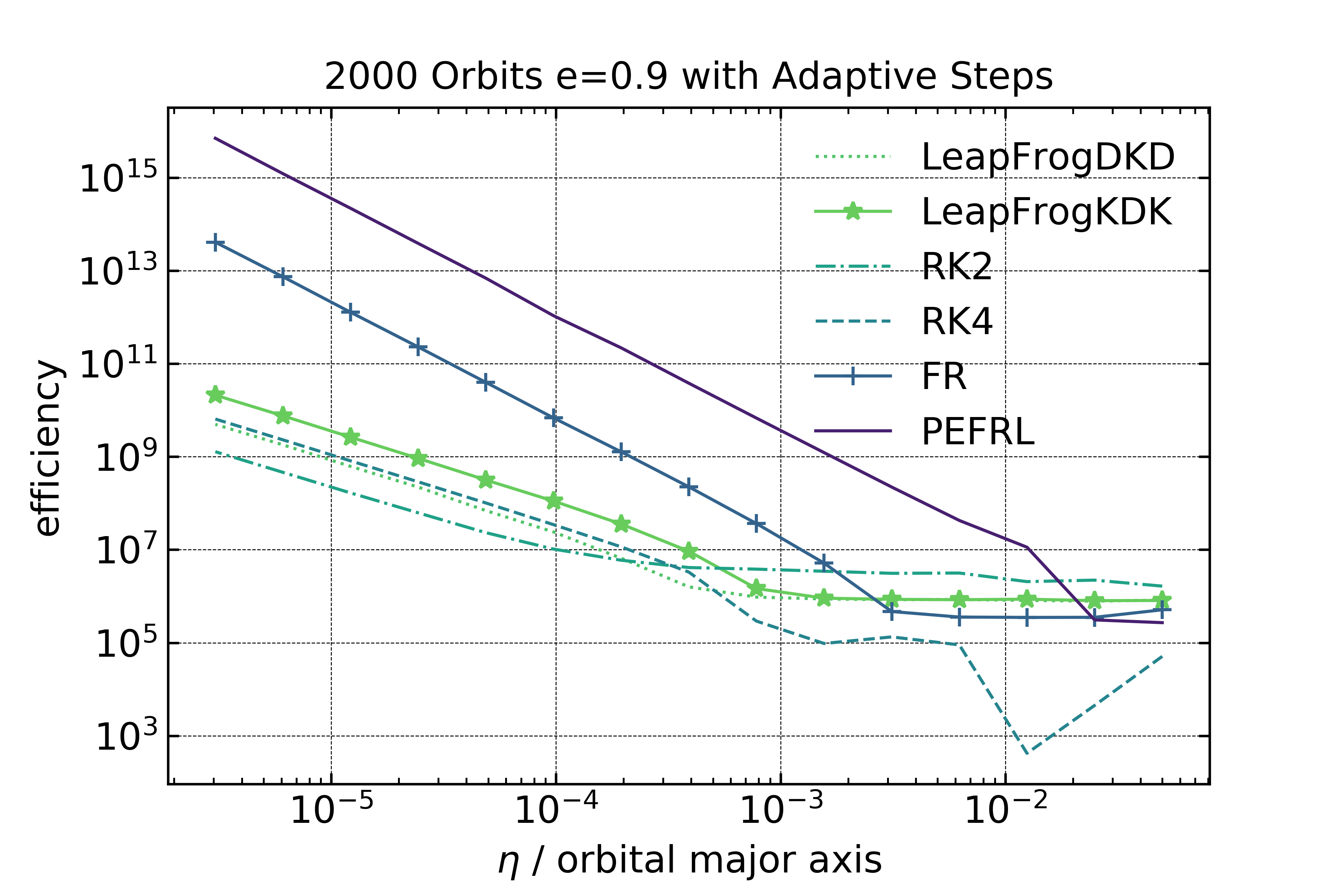}\\
\includegraphics[width=\columnwidth]{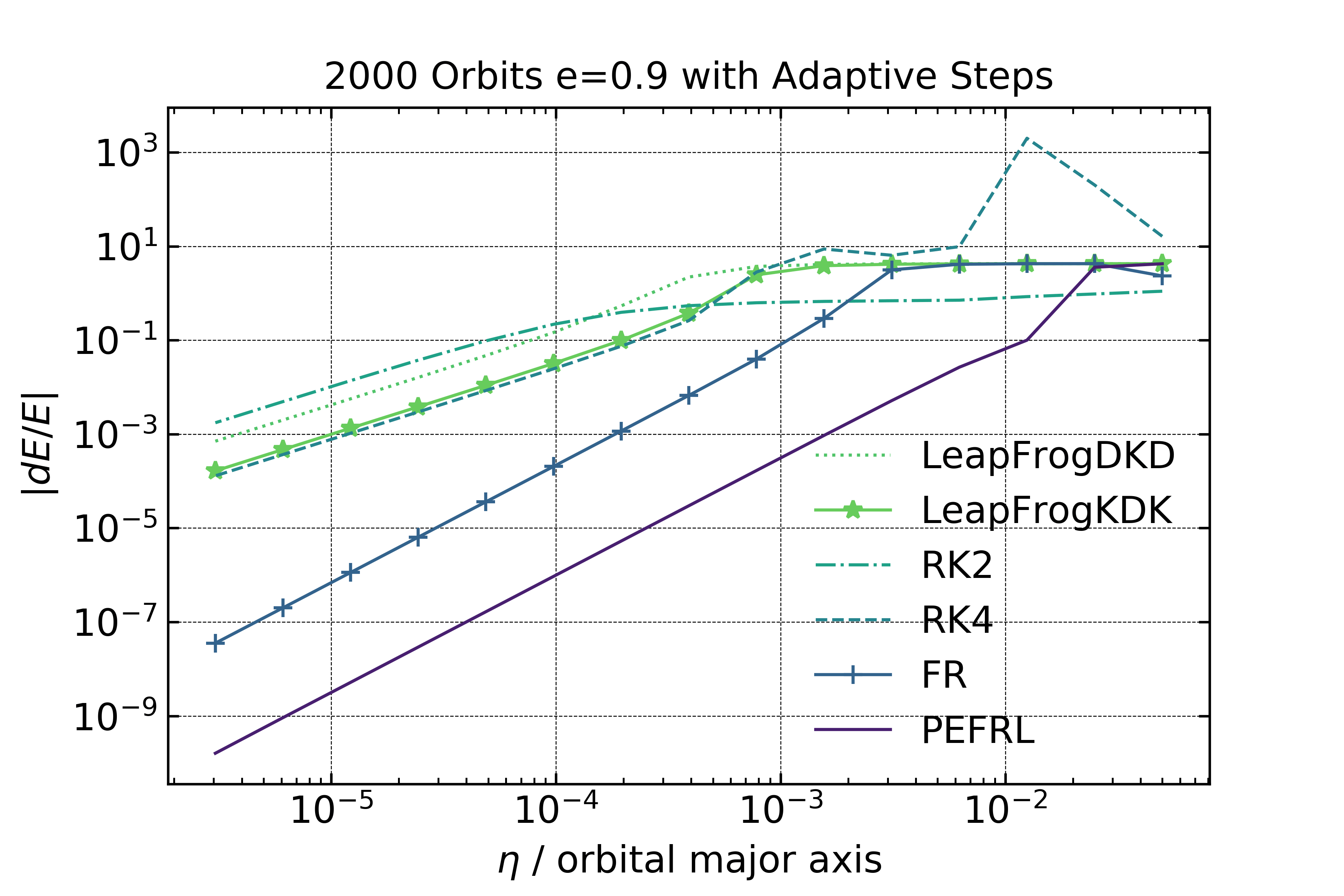}\\
\includegraphics[width=\columnwidth]{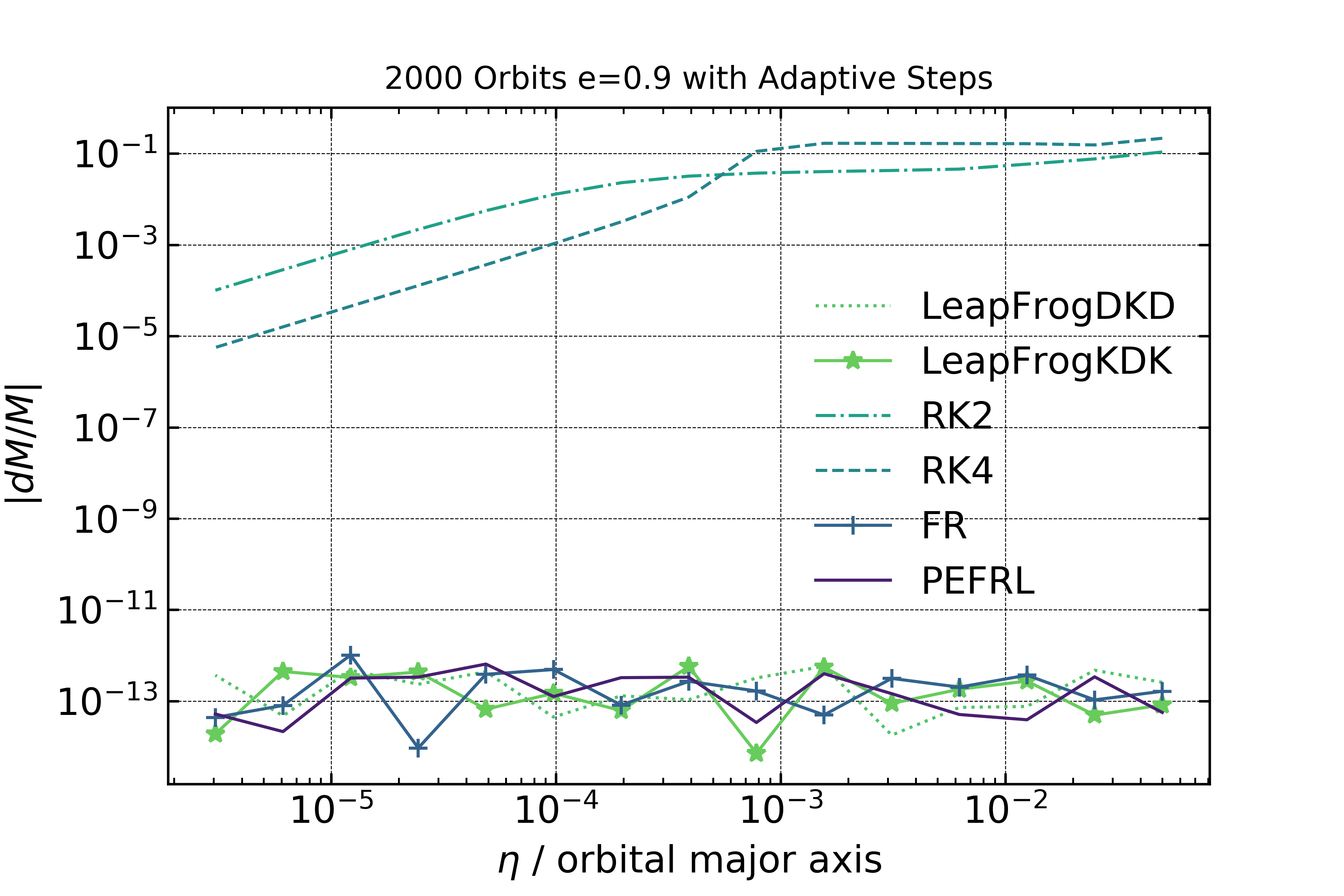}
\caption{{\it Top panel:} The total time-steps versus the precision 
  parameter$\eta$, which shows the efficiency of the algorithms for
  the different values of the precision parameter. 
  {\it Middle panel:} The energy fluctuations as a  function $\eta$.
  {\it Bottom panel:} The angular momentum fluctuations as a function $\eta$.}
\label{fig:symp}
\end{figure}

Figure \ref{fig:symp} compares the performance of several algorithms in computing long period Keplerian orbits with the adaptive time-step method. Leapfrog and RK2 are second order algorithms. RK4, FR and PEFRL are fourth order algorithms. Leapfrog, FR and PEFRL are symplectic algorithms. RK2 and RK4 are non-symplectic algorithms.

\begin{thebibliography}{99}

\bibitem[\protect\citeauthoryear{Alexander}{2005}]{Alexander} Alexander T., 2005, PhR, 419, 65

\bibitem[\protect\citeauthoryear{Anderson, Storch, \& Lai}{2016}]{Anderson} Anderson K.~R., Storch N.~I., Lai D., 2016, MNRAS, 456, 3671

\bibitem[\protect\citeauthoryear{Antonini et al.}{2010}]{Antonini 09} Antonini F., Faber J., Gualandris A., Merritt D., 2010, ApJ, 713, 90
{ 
\bibitem[\protect\citeauthoryear{Antonini, Murray, \& Mikkola}{2014}]{Antonini2014} Antonini F., Murray N., Mikkola S., 2014, ApJ, 781, 45
}
\bibitem[\protect\citeauthoryear{Antonini, Murray \& Mikkola}{2014}]{AMM} Antonini F., Murray N., Mikkola S., 2014, ApJ, 781, 45

\bibitem[\protect\citeauthoryear{Blaes et al.}{2002}]{Blaes} Blaes O., Lee M. H., Socrates A., 2002, ApJ, 578, 775

\bibitem[\protect\citeauthoryear{Brown et al.}{2010}]{brown10} Brown W.~R., Anderson J., Gnedin O.~Y., Bond H.~E., Geller M.~J., Kenyon S.~J., Livio M. 2010, ApJL, 719, L23
{ 
\bibitem[\protect\citeauthoryear{Capuzzo-Dolcetta \& Fragione}{2015}]{Capuzzo2015} Capuzzo-Dolcetta R., Fragione G., 2015, MNRAS, 454, 2677

\bibitem[\protect\citeauthoryear{Fragione \& Capuzzo-Dolcetta}{2016}]{Capuzzo2016} Fragione G., Capuzzo-Dolcetta R., 2016, MNRAS, 458, 2596

\bibitem[\protect\citeauthoryear{Fragione, Capuzzo-Dolcetta, \& Kroupa}{2017}]{Capuzzo2017} Fragione G., Capuzzo-Dolcetta R., Kroupa P., 2017, MNRAS, 467, 451

}
\bibitem[\protect\citeauthoryear{Chen et al.}{2009}]{liufukun} Chen X., Madau P., Sesana A., Liu F.~K., 2009, ApJ, 697, L149

\bibitem[\protect\citeauthoryear{Eggleton \& Kiseleva\---Eggleton}{2001}]{Eggleton} Eggleton P. P. \& Kiseleva\---Eggleton L., 2001, ApJ, 562, 1012

\bibitem[\protect\citeauthoryear{Fabrycky \& Tremaine}{2007}]{FT} Fabrycky D. C., Tremaine S. 2007 ApJ 669 1298

{ 
\bibitem[\protect\citeauthoryear{Fragione \& Loeb}{2017}]{Fragione2017} Fragione G., Loeb A., 2017, NewA, 55, 32

\bibitem[\protect\citeauthoryear{Fragione, Ginsburg, \& Kocsis}{2017}]{Fragione2017} Fragione G., Ginsburg I., Kocsis B., 2017, arXiv, arXiv:1711.00483
}
\bibitem[\protect\citeauthoryear{Georgiev et al.}{2016}]{georgiev16} Georgiev I., B\"oker T., Leigh N. W. C., L\"utzgendorf N., Neumayer N. 2016, MNRAS, 457, 2122 

\bibitem[\protect\citeauthoryear{Gezari et al.}{2012}]{Gezari} Gezari S., et al., 2012, Nature, 485, 217

{ 
\bibitem[\protect\citeauthoryear{Gillessen et al.}{2017}]{Gillessen2017} Gillessen S., et al., 2017, ApJ, 837, 30 
}
\bibitem[\protect\citeauthoryear{Ginsburg \& Loeb}{2007}]{Ginsburg} Ginsburg I., Loeb A., 2007, MNRAS, 376, 492

\bibitem[\protect\citeauthoryear{Gualandris, Portegies Zwart \& Sipior}{2005}]{gualandris05} Gualandris A., Portegies Zwart S., Sipior M. S., 2005, MNRAS, 363, 223

{ 
\bibitem[\protect\citeauthoryear{Gualandris \& Merritt}{2009}]{Gualandris2009} Gualandris A., Merritt D., 2009, ApJ, 705, 361 
}
\bibitem[\protect\citeauthoryear{Hansen, Kawaler, \& Trimble}{2004}]{Hansen} Hansen C.~J., Kawaler S.~D., Trimble V., 2004, sipp.book

\bibitem[\protect\citeauthoryear{Hills}{1975}]{Hills T} Hills J.~G., 1975, Nature, 254, 295

\bibitem[\protect\citeauthoryear{Hills}{1988}]{Hills H} Hills J.~G., 1988, Nature, 331, 687

\bibitem[\protect\citeauthoryear{Holman, Touma,\& Tremaine}{1997}]{Holman} Holman M., Touma J., Tremain S., 1997, Nature, 386, 254

\bibitem[\protect\citeauthoryear{Innanen et al.}{1997}]{Innanen} Innanen K. A., Zheng J. Q., Mikkola S., Valtonen M. J., 1997, AJ, 113, 1915

\bibitem[\protect\citeauthoryear{Katz \& Dong}{2012}]{Katz} Katz B., Dong S., 2012, preprint (arXiv:1211.4584)
{ 
\bibitem[\protect\citeauthoryear{Kenyon et al.}{2014}]{Kenyon2014} Kenyon S.~J., Bromley B.~C., Brown W.~R., Geller M.~J., 2014, ApJ, 793, 122 
}
\bibitem[\protect\citeauthoryear{Kormendy \& Ho}{2013}]{Ho} Kormendy J., Ho L.~C., 2013, ARA\&A, 51, 511

\bibitem[\protect\citeauthoryear{Kozai}{1962}]{Kozai} Kozai Y., 1962, AJ, 67, 591

\bibitem[\protect\citeauthoryear{Kushnir et al.}{2013}]{Kushnir} Kushnir D., Katz B., Dong S., Livne E., Fern{\'a}ndez R., 2013, ApJ, 778, L37

\bibitem[\protect\citeauthoryear{Kupi, Amaro-Seoane, \& Spurzem}{2006}]{PN} Kupi G., Amaro-Seoane P., Spurzem R., 2006, MNRAS, 371, L45 

\bibitem[\protect\citeauthoryear{Leigh \& Geller}{2013}]{leigh13} Leigh N.~W.~C., Geller A.~M. 2013, MNRAS, 432, 2474

\bibitem[\protect\citeauthoryear{Leigh et al.}{2016a}]{leigh16a} Leigh N.~W.~C., Antonini F., Stone N.~C., Shara M.~M., Merritt D. 2016, MNRAS, 463, 1605

\bibitem[\protect\citeauthoryear{Leigh et al.}{2016b}]{Leigh2016} Leigh N.~W.~C., Stone N.~C., Geller A.~M., Shara M.~M., Muddu H., Solano-Oropeza D., Thomas Y., 2016, MNRAS, 463, 3311

\bibitem[\protect\citeauthoryear{Li et al.}{2015}]{Li TDE} Li G., Naoz S., Kocsis B., Loeb A., 2015, MNRAS, 451, 1341

\bibitem[\protect\citeauthoryear{Lidov}{1962}]{Lidov} Lidov M. L., 1962, Planet. Space Sci., 9, 719

\bibitem[\protect\citeauthoryear{Lu, Yu, \& Lin}{2007}]{2007ApJ...666L..89L} Lu Y., Yu Q., Lin D.~N.~C., 2007, ApJ, 666, L89 
{ 
\bibitem[\protect\citeauthoryear{Liu, Wang, \& Yuan}{2017}]{Liu2017} Liu B., Wang Y.-H., Yuan Y.-F., 2017, MNRAS, 466, 3376 
}
\bibitem[\protect\citeauthoryear{Liu, Lai, \& Yuan}{2015}]{Resonance} Liu B., Lai D., Yuan Y.-F., 2015, PhRvD, 92, 124048


\bibitem[\protect\citeauthoryear{Mandel \& Levin}{2015}]{DoubleTDE} Mandel I., Levin Y., 2015, ApJ, 805, L4
{
\bibitem[\protect\citeauthoryear{Marchetti et al.}{2017}]{Marchetti2017} Marchetti T., Contigiani O., Rossi E.~M., Albert J.~G., Brown A.~G.~A., Sesana A., 2017, arXiv, arXiv:1711.11397 
}
\bibitem[\protect\citeauthoryear{Martins et al.}{2006}]{martins06}  Martins F., et al. 2006, ApJ, 649, L103


\bibitem[\protect\citeauthoryear{Merritt}{2013}]{merritt13} Merritt D. 2013, Dynamics and Evolution of Galactic Nuclei (Princeton:  Princeton University Press)

\bibitem[\protect\citeauthoryear{Miller \& Hamilton}{2002}]{MH} Miller M. C., Hamilton D.P., 2002, ApJ, 576, 894


\bibitem[\protect\citeauthoryear{Naoz et al.}{2013a}]{Smadar 2013a} Naoz S., Kocsis B., Loeb A., Yunes N., 2013a, ApJ, 773, 187

{ 
\bibitem[\protect\citeauthoryear{Naoz}{2016}]{Naoz2016} Naoz S., 2016, ARA\&A, 54, 441 

\bibitem[\protect\citeauthoryear{Perets}{2009}]{perets09} Perets H.~B., 2009, ApJ, 698, 1330 
}

\bibitem[\protect\citeauthoryear{Perets \& Subr}{2012}]{perets12} Perets H.~B., Subr L. 2012, ApJ, 751, 133 

\bibitem[\protect\citeauthoryear{Phinney}{1989}]{Phinney} Phinney E.~S., 1989, IAUS, 136, 543

\bibitem[\protect\citeauthoryear{Prodan, Murray, \& Thompson}{2013}]{PMT} Prodan S., Murray N., Thompson T.~A., 2013, preprint (arXiv:1305.2191)
{ 
\bibitem[\protect\citeauthoryear{Prodan, Antonini, \& Perets}{2015}]{Prodan2015} Prodan S., Antonini F., Perets H.~B., 2015, ApJ, 799, 118 
}
\bibitem[\protect\citeauthoryear{Rees}{1988}]{Rees} Rees M.~J., 1988, Nature, 333, 523

{ 
\bibitem[\protect\citeauthoryear{Ryu, Leigh, \& Perna}{2017}]{Ryu17a} Ryu T., Leigh N.~W.~C., Perna R., 2017, MNRAS, 470, 3049 


\bibitem[\protect\citeauthoryear{Ryu, Leigh, \& Perna}{2017}]{Ryu17b} Ryu T., Leigh N.~W.~C., Perna R., 2017, MNRAS, 467, 4447

\bibitem[\protect\citeauthoryear{Sesana, Madau, \& Haardt}{2009}]{Sesana09} Sesana A., Madau P., Haardt F., 2009, MNRAS, 392, L31
}
\bibitem[\protect\citeauthoryear{Sesana, Haardt, \& Madau}{2008}]{Sesana08} Sesana A., Haardt F., Madau P., 2008, ApJ, 686, 432-447 


\bibitem[\protect\citeauthoryear{Sesana, Haardt, \& Madau}{2007}]{Sesana07} Sesana A., Haardt F., Madau P., 2007, ApJ, 660, 546 


\bibitem[\protect\citeauthoryear{Sesana, Haardt, \& Madau}{2006}]{Sesana06} Sesana A., Haardt F., Madau P., 2006, ApJ, 651, 392 
\bibitem[\protect\citeauthoryear{Soffel}{1989}]{Soffel1989} Soffel M.~H., 1989, S\&T, 78, 382

\bibitem[\protect\citeauthoryear{Storch, Anderson, \& Lai}{2014}]{DongLai S} Storch N.~I., Anderson K.~R., Lai D., 2014, Sci, 345, 1317


\bibitem[\protect\citeauthoryear{Springel}{2005}]{Springel05} Springel V., 2005, MNRAS, 364, 1105 

\bibitem[\protect\citeauthoryear{Thompson}{2011}]{Thompson} Thompson T.~A., 2011, ApJ, 741, 82

\bibitem[\protect\citeauthoryear{Wen}{2003}]{Wen} Wen L., 2003, ApJ, 598, 419

\bibitem[\protect\citeauthoryear{Will}{2014a}]{Clifford1} Will C.~M., 2014, PhRvD, 89, 044043

\bibitem[\protect\citeauthoryear{Will}{2014b}]{Clifford2} Will C.~M., 2014, CQGra, 31, 244001


\bibitem[\protect\citeauthoryear{Wu \& Murray}{2003}]{WM} Wu Y., Murray N., 2003, ApJ, 589, 605

\bibitem[\protect\citeauthoryear{Yu \& Tremaine}{2003}]{Yuqingjuan} Yu Q., Tremaine S., 2003, ApJ, 599, 1129
\end{thebibliography}
\end{document}